\documentclass[english, 11pt,a4paper]{article}
\pdfoutput=1
\usepackage{hyperref}
\usepackage{graphicx,color}
\RequirePackage{amsthm,amsmath,natbib}
\usepackage{epstopdf}
\usepackage{amsmath,amsfonts,bbm}
\setcitestyle{square,numbers}
\let\OLDthebibliography\thebibliography
\renewcommand\thebibliography[1]{
  \OLDthebibliography{#1}
  \setlength{\parskip}{0pt}
  \setlength{\itemsep}{0pt plus 0.3ex}
}

\usepackage{amsmath, amsthm,mathtools, amssymb,upgreek,slashed}

\begin{document}


\font\msytw=msbm10 scaled\magstep1
\font\msytww=msbm7 scaled\magstep1
\font\msytwww=msbm5 scaled\magstep1
\font\cs=cmcsc10
\font\ottorm=cmr8

\let\a=\alpha \let\b=\beta  \let\g=\gamma  \let\d=\delta \let\e=\varepsilon
\let\z=\zeta  \let\h=\eta   \let\th=\theta \let\k=\kappa \let\l=\lambda
\let\m=\mu    \let\n=\nu    \let\x=\xi     \let\p=\pi    \let\r=r
\let\s=\sigma \let\t=\tau   \let\f=\varphi \let\ph=\varphi\let\c=\chi
\let\ps=\psi  \let\y=\upsilon \let\o=\omega\let\si=\varsigma
\let\G=\Gamma \let\D=\Delta  \let\Th=\Theta\let\L=\Lambda \let\X=\Xi
\let\P=\Pi    \let\Si=\Sigma \let\F=\Phi    \let\Ps=\Psi
\let\O=\Omega \let\Y=\Upsilon

\def\ins#1#2#3{\vbox to0pt{\kern-#2 \hbox{\kern#1 #3}\vss}\nointerlineskip}

\newdimen\xshift \newdimen\xwidth \newdimen\yshift

\def\vdd{{\vec d}}\def\vee{{\vec e}}\def\vkk{{\bk}}\def\vii{{\vec i}}
\def\vmm{{\vec m}}\def\vnn{{\vec n}}\def\vpp{{\vec p}}\def\vqq{{\vec q}}
\def\vxxi{{\vec \xi}}\def\vrr{{\vec r}}\def\vtt{{\vec t}}
\def\vuu{{\vec u}}\def\vvv{{\vec v}}
\def\vxx{{\xx}}\def\vyy{{\vec y}}\def\vzz{{\vec z}}
\def\un{{\underline n}} \def\ux{{\underline x}} \def\uk{{\underline k}}
\def\xxx{{\underline\xx}}\def\vxx{{\xx}} \def\vxxx{{\underline\vxx}}
\def\kkk{{\underline\kk}} \def\vkkk{{\underline\vkk}}
\def\bO{{\bf O}}\def\rr{{\bf r}} \def\bk{{\bf k}}  \def\bp{{\bf p}}
 \def\bP{{\bf P}}\def\bl{{\bf l}} \def\bF{{\bf F}}
\def\ss{{\underline \sigma}}\def\oo{{\underline \omega}}

\def\PPP{{\cal P}}\def\EE{{\cal E}}\def\cF{{\cal F}}
\def\mF{{\mathfrak{F}}}
\def\MM{{\cal M}} \def\VV{{\cal V}}\def\cB{{\cal B}}\def\cA{{\cal A}}\def\cI{{\cal I}}
\def\bV{{\bf V}_n}
\def\cE{{\cal E}}\def\si{{\sigma}}\def\ep{{\epsilon}} \def\cD{{\cal D}}\def\cG{{\cal G}}
\def\CC{{\cal C}}\def\FF{{\cal F}} \def\FFF{{\cal F}}\def\cJ{{\cal J}}
\def\cF{{\cal F}} \def\cT{{\cal T}}\def\cS{{\cal S}}\def\cQ{{\cal Q}}
\def\HHH{{\cal H}}\def\WW{{\cal W}}\def\cP{{\cal P}}
\def\TT{{\cal T}}\def\NN{{\cal N}} \def\BBB{{\cal B}}\def\III{{\cal I}}
\def\RR{{\cal R}}\def\cL{{\cal L}} \def\JJ{{\cal J}} \def\OO{{\cal O}}
\def\DD{{\cal D}}\def\AAA{{\cal A}}\def\GG{{\cal G}} \def\SS{{\cal S}}
\def\KK{{\cal K}}\def\UU{{\cal U}} \def\QQ{{\cal Q}} \def\XXX{{\cal X}}

\def\hh{{\bf h}} \def\HH{{\bf H}} \def\AA{{\bf A}} \def\qq{{\bf q}}
\def\bG{{\bf G}}
\def\BB{{\bf B}} \def\XX{{\bf X}} \def\PP{{\bf P}} \def\bP{{\bf P}} 
\def\bp{{\bf p}} \def\bq{{\bf q}}
\def\vv{{\bf v}} \def\xx{{\bf x}} \def\yy{{\bf y}} \def\zz{{\bf z}}
\def\dd{{\bf d}} \def\eee{{\bf e}} \def\bn{{\bf n}}
\def\aaa{{\bf a}}\def\bb{{\bf b}}\def\hhh{{\bf h}}\def\II{{\bf I}}
\def\ii{{\bf i}}\def\jj{{\bf j}}\def\kk{{\bf k}}\def\bS{{\bf S}}
\def\mm{{\bf m}}\def\Vn{{\bf n}}\def\uu{{\bf u}}\def\tt{{\bf t}}
\def\B{\hbox{\msytw B}}
\def\RRR{\hbox{\msytw R}} \def\rrrr{\hbox{\msytww R}}\def\mI{\hbox{\msytw I}} 
\def\rrr{\hbox{\msytwww R}} \def\CCC{\hbox{\msytw C}}\def\EEE{\hbox{\msytw E}}
\def\cccc{\hbox{\msytww C}} \def\ccc{\hbox{\msytwww C}}
\def\MMM{\hbox{\euftw M}}\font\euftw=eufm10 scaled\magstep1%
\def\NNN{\hbox{\msytw N}} \def\nnnn{\hbox{\msytww N}}
\def\nnn{\hbox{\msytwww N}} \def\ZZZ{\hbox{\msytw Z}}\def\QQQ{\hbox{\msytw Q}}
\def\zzzz{\hbox{\msytww Z}} \def\zzz{\hbox{\msytwww Z}}
\def\SSS{{\bf S}}
\def\SSSS{\hbox{\euftwww S}}
\def\1{\hbox{\msytw 1}}
\newcommand{\mR}{{\msytw R}}
\def\virg{\quad,\quad}


\def\\{\hfill\break}
\def\={:=}
\let\io=\infty
\let\0=\noindent\def\pagina{{\vfill\eject}}
\def\media#1{{\langle#1\rangle}}
\let\dpr=\partial
\def\sign{{\rm sign}}
\def\const{{\rm const}}
\def\tende#1{\,\vtop{\ialign{##\crcr\rightarrowfill\crcr\noalign{\kern-1pt
    \nointerlineskip} \hskip3.pt${\scriptstyle #1}$\hskip3.pt\crcr}}\,}
\def\otto{\,{\kern-1.truept\leftarrow\kern-5.truept\to\kern-1.truept}\,}
\def\defin{{\buildrel def\over=}}
\def\wt{\widetilde}
\def\wh{\widehat}
\def\to{\rightarrow}
\def\ra{\right\rangle}
\def\qed{\hfill\raise1pt\hbox{\vrule height5pt width5pt depth0pt}}
\def\Val{{\rm Val}}
\def\ul#1{{\underline#1}}
\def\lis{\overline}
\def\V#1{{\bf#1}}
\def\be{\begin{equation}}
\def\ee{\end{equation}}
\def\bea{\begin{eqnarray}}
\def\eea{\end{eqnarray}}
\def\bd{\begin{definition}}
\def\ed{\end{definition}}

\def\nn{\nonumber}
\def\pref#1{(\ref{#1})}
\def\ie{{\it i.e.}}
\def\cC{{\cal C}}
\def\lb{\label}
\def\eg{{\it e.g.}}
\def\sl{{\displaystyle{\not}}}
\def\Tr{\mathrm{Tr}}
\def\BBBB{\hbox{\msytw B}}
\def\bbb{\hbox{\msytww B}}
\def\TTT{\hbox{\msytw T}}
\def\d{\delta}
\def\bT{{\bf T}}
\def\mod{{\rm mod}}
\def\der{{\rm d}}
\def\bs{\backslash}

\newtheorem{corollary}{Corollary}[section]
\newtheorem{lemma}{Lemma}[section]
\newtheorem{conjecture}{Conjecture}[section]
\newtheorem{notation}{Notation}[section]
\newtheorem{example}{Example}[section]

\newtheorem{remark}{Remark}[section]
\newtheorem{definition}{Definition}[section]
\newtheorem{theorem}{Theorem}[section]
\newtheorem{proposition}{Proposition}[section]
\newtheorem{oss}{Remark}

\renewcommand{\thesubsection}{\arabic{section}.\arabic{subsection}}

\renewcommand{\theequation}{\thesection.\arabic{equation}}

\title{{\bf Phase Transitions in the Hubbard Model on the Square Lattice\\
}}

\author{Zhituo Wang\\
Institute for Advanced Study in Mathematics, \\Harbin Institute of Technology\\
Email: wzht@hit.edu.cn}

\maketitle

\begin{abstract}
We study the low temperature properties of the two-dimensional weakly interacting Hubbard model on $\ZZZ^2$ with renormalized chemical potential $\mu=2-\mu_0$, $\mu_0=10^{-10}$ being fixed. The Fermi surface is close to a perfect square without van Hove singularities. Using fermionic cluster expansions and rigorous renormalization group analysis, we prove that the perturbation series for the two-point Schwinger function is analytic in the coupling constant $\l$ in the domain $\l\in\RR_T=\{\l\in\RRR,\vert\lambda\log^2(\mu_0T/C_1)|\le C_2\}$ for any fixed temperature $T>0$, suggesting that there is a phase transition with critical temperature $T_c= \frac{C_1}{\m_0}\exp{(-C^{1/2}_2|\lambda|^{-1/2})}$. 
Here $C_1, C_2$ are positive constants independent of $T$ and $\l$. We also establish the optimal upper bound for the self-energy function and prove that the second derivative of the momentum space self-energy function w.r.t. the external momentum is not uniformly bounded, suggesting that this model is {\it not} a Fermi liquid in the mathematically precise sense of Salmhofer. This result can be viewed as a first step towards a rigorous study of the crossover phenomenon between the Fermi liquid and the non-Fermi liquid.
\end{abstract}

\vskip.5cm

\renewcommand{\thesection}{\arabic{section}}
\section{Introduction}
The Hubbard model \cite{hubb} is a prototypical model in mathematical physics. It describes the interacting electrons on the lattice, taking into account the quantum mechanical hopping of the electrons and the screened Coulomb interactions between them, through a density-density interaction.
Despite its simple definition, this model contains all necessary terms to describe the basic properties of correlated electron systems and exhibits various interesting phenomena, including the Fermi liquid behaviors, the metal-insulator transitions and finally the high-$T_c$ superconductivity \cite{lieb}. Therefore, it is considered as {\it the standard model} of correlated electron systems, playing the same role as the Ising model in classical statistical mechanics. The $1d$-Hubbard model is well understood and many rigorous results have been obtained using the Bethe ansatz \cite{hubb1d} and the renormalization group analysis \cite{M2}. However, we are still far from a rigorous understanding of many of its properties in $2$ or higher dimensions. Whether the low temperature equilibrium state (also called the ground state) is a Fermi liquid or not in $2d$ Hubbard model is rather controversial. The nature of the ground state of the Hubbard model depends crucially on the geometry of the non-interacting Fermi surface $\cF_0$, which depends on the spatial lattice $\L$ and the chemical potential $\mu$. In the square Hubbard model, $\L=\ZZZ^2$, crossover phenomenon happens for different values of $\mu$: For $0<\mu\le1$, the Fermi surface is close to a circle, and the ground state is proved to be a Fermi liquid for temperature greater than an exponential one \cite{BGM2}. For $\mu=2$, in which the system is at half-filling, the non-interacting Fermi surface is a square and the ground state is proved {\it not} to be a Fermi liquid, in the mathematically precise sense of Salmhofer \cite{salm, Riv04, AMR1}. Similar phenomenon happens in the $2d$ honeycomb-Hubbard model \cite{GM, RWang1}, which is an important model for studying  the {\it Graphene} \cite{N, feff2, feff3}. When $\mu=0$, the Fermi surface composes of a set of points \cite{N, feff3} and this model displays Fermi liquid behavior \cite{GM}. But when the renormalized chemical potential is equal to $1$, the Fermi surface becomes a set of exact triangles, and the ground state is not a Fermi liquid \cite{RWang1}. Founding a mathematical theory of the crossover phenomenon between the Fermi liquid and the non-Fermi liquid  is still an open problem in mathematical physics. 

One way of attacking this problem is to consider transitions in a two-dimensional "model space" of
the Hubbard models on a fix lattice with all possible values of $\l$ and $\mu$. Fixing a sufficiently small $\l$, then the different ground states can be considered as different phases in a one-dimensional subspace labeled by $\mu$. The crossover phenomenon can be studied by deforming the chemical potential $\mu$ starting from a given state, which can be a Fermi liquid or not, and determine at which critical value $\mu_c$ does this crossover phenomenon happen. In practice this picture can be more complicated as the parameters $\l$ and $\mu$ are moving in the model space when the temperature $T$ changes, due to the many-body interactions. The present paper can be considered as a first step in this approach. 

In this paper we consider the 2d square Hubbard model with {\it renormalized} chemical potential $\mu=2-\mu_0$, $\mu_0=10^{-10}$ fixed. In this setting the non-interacting Fermi surface is a convex curve close to a square and van Hove singularities disappear. The main results of the present paper is the following: We establish the power counting theorem for the $2p$-point Schwinger function, $p\ge1$ and prove that the perturbation series for the two-point many-fermion systems as well as the self-energy function have positive radius of convergence when the temperature $T$ is greater than the critical value $T_c\sim\frac{1}{\m_0^3}\exp{(-C^{1/2}|\lambda|^{-1/2})}$, where $|\lambda|\ll1$ is the bare coupling constant, $C$ is a positive constant depending on the physical parameters of the model such as the electron mass, the lattice structure, etc., but not on the temperature. The optimal non-perturbative upper bounds for the self-energy function and its second derivatives w.r.t. the external momenta have been established. The fact that the analytic domain (see Theorem \ref{mainthm}) of the two-point Schwinger functions is smaller than that required by Salmhofer's criterion, and that the second derivative of the self-energy is not uniformly bounded for $T\rightarrow0$ (see Theorem \ref{mainb} and Remark \ref{rmkb}) suggest that the ground state is {\it not} a Fermi liquid. Notice that it has been conjectured in \cite{vieri1} that the ground state of this model is a non-Fermi liquid when $\mu$ is slightly smaller than $2$, but no rigorous proof existed before. 

We believe that this work is important since it addresses the Hubbard model close to half-filling, which is of current interest in the physics community, and mathematically it provides the first rigorous results on the non-Fermi liquid behavior on a many-fermion system without van Hove singularity.

The main results of this paper will be proved with the fermionic cluster expansion \cite{BK, AR}, sector analysis \cite{FMRT, DR2, BGM2, wang20} and renormalization group (RG) analysis \cite{BG, FT, M2}. Since the Fermi surface is not a square, sectors have been defined differently than \cite{Riv04} (see Definition \ref{defsec}) and the sector analysis is also different. One major difficulty in the proof is that the non-interacting Fermi surface is deformed by interaction, and the resulting interacting Fermi surface $\cF$ is moving. This shift of Fermi surface may cause divergence of the coefficients in the naive perturbation expansion. In order to solve this problem, we introduce counter-terms to the interaction potential, in such a way that the interacting Fermi surface for the {\it new model} is fixed and coincides with $\cF_0$. The inversion problem, which concerns the existence and uniqueness of the counter-term given a bare dispersion is not addressed in this paper.

The organization of this paper is as follows. In Section $1$ we introduce the model and main results. In Section $2$ we introduce the fermionic functional integral representation for the Schwinger functions and the multi-scale analysis. Section $3$ is devoted to the convergent perturbation expansion of the Schwinger functions. Section $4$ and $5$ are about the analyticity of the Schwinger functions. In Section $6$ we study the upper bounds of the self-energy function. The appendix is devoted to the geometry of the Fermi surface.

%


\subsection{The Model and Main results}
Let $\L=\ZZZ^2$ be a square lattice and $\L_L=\ZZZ^2/L\ZZZ^2$ be a finite sub-lattice with side $L\in\ZZZ_+$ and metric $d_L:=|\xx-\yy|_{\L_L}=\min_{(n_1,n_2)\in\ZZZ^2}\vert\xx-\yy+n_1{\eee_1}+n_2\eee_2\vert$. The sites on the lattice are labeled by $(n_1,n_2)\in\ZZZ^2$, with $-L/2\le n_1,n_2\le L-1/2$ and $\eee_1=(1,0)$, $\eee_2=(0,1)$ are the unit vectors on $\ZZZ^2$. The Fock space for the many-fermion system is constructed as follows:
let $\HHH_L=\CCC^{L^2}\otimes\CCC^2$ be a single-particle Hilbert space of functions {$\Psi_{\xx, \t}:\L_L\times\{\uparrow,\downarrow\}\rightarrow\CCC$}, in which $\t\in\{\uparrow,\downarrow\}$ labels the spin index of the quasi-particle and $\xx\in \L_L$ labels the lattice point. The
normalization condition is $\Vert\Psi\Vert^2_2=\sum_{\xx,\t}\vert\Psi_{\xx,\t}\vert^2=1$. The Fermionic Fock space $\bF_{L}$ is defined as:
\be
{\bF_{L}}=\CCC\oplus\bigoplus_{N=1}^{L^2}\bF_{L}^{(N)},\quad \bF_{L}^{(N)}=\bigwedge^N \HHH_L,
\ee
where $\bigwedge^N \HHH_L$ is the $N$-th anti-symmetric tensor product of $\HHH_L$. The vector
$\Omega:=(1,0,0,\cdots)\in\bF_L$ is called the vacuum. Let $\xi_i=(\xx_i,\t_i)$, $i=1,\cdots,N$. 
For $\Psi=(\Psi^{(0)},\Psi^{(1)},\Psi^{(2)},\cdots)\in\bF_L$ and $\psi\in \HHH_L$, define the fermionic creation operators $\{{\bf a}_{\xx,\t}^+(\psi)\}$ and annihilation operators $\{{\bf a}_{\xx,\t}^-(\psi)\}$ by their actions on ${\bF_{L}}$ ({see eg. \cite{BR1}, Page 10, Example 5.2.1}):
\bea
&&({\bf a}^+_{\xx,\t}(\psi)\Psi)^{(N)}(\xi_1,\cdots, \xi_N)\nn\\
&&\quad\quad\quad:=\frac{1}{\sqrt{N}}\sum_{j=1}^N(-1)^{j-1}\psi(\xi) \Psi^{(N-1)}(\xi_1,\cdots ,\xi_{j-1},\xi_{j+1},\cdots,\xi_{N}),\\
&&({\bf a}^-_{\xx,\t}(\psi)\Psi)^{(N)}(\xi_1,\cdots, \xi_N):= \sqrt{N+1}\ \sum_{\xi\in \L_L\times\{\uparrow,\downarrow\}}\bar\psi(\xi)\Psi^{(N+1)}(\xi;\ \xi_1,\cdots,\xi_{n}).
\eea 
It is also convenient to define the fermionic field operators $\{{\bf a}_{\xx,\t}^\pm\}$, which are operator valued distributions on $\L_L\times\{\uparrow,\downarrow\}$, through the relation
\be
{\bf a}_{\xx,\t}^-(\psi)=\sum_{\xi\in\L_L\times\{\uparrow,\downarrow\}}\bar\psi(\xi){\bf a}_{\xx,\t}^-,\ {\bf a}_{\xx,\t}^+(\psi)=\sum_{\xi\in \L_L\times\{\uparrow,\downarrow\}}\psi(\xi){\bf a}_{\xx,\t}^+.
\ee
They satisfy the canonical anti-commutation relations (CAR) \cite{BR1}: \be
\{{\bf a}^+_{\xx,\t},{\bf a}^-_{\xx',\t'}\}=\delta_{\xx,\xx'}\delta_{\tau,\tau'},\ \{{\bf a}^+_{\xx,\t},{\bf a}^+_{\xx',\t'}\}=0=\{{\bf a}^-_{\xx,\t},{\bf a}^-_{\xx',\t'}\},
\ee
in which $\delta$ is the Kronecker delta-function. We impose the periodic boundary conditions on these fermionic field operators: ${\bf a}^\pm_{\xx+n_1L+n_2L,\t}={\bf a}^\pm_{\xx,\t}$, $\forall\xx\in\L_L$. 

The second quantized grand-canonical Hamiltonian on $\L_L$ is defined by:
\bea
H_{L}&=&t\sum_{\xx\in\L_L}\sum_{\t=\uparrow\downarrow}
{\bf a}^+_{\xx,\t}(-\frac{\Delta}{2}) {\bf a}^-_{\xx, \t} -\mu\sum_{\xx\in\L_L}\sum_{\t=\uparrow\downarrow}
{\bf a}^+_{\xx,\t}{\bf a}^-_{\xx, \t}+\lambda\sum_{\xx\in\L_L} n_{\xx,\uparrow}n_{\xx,\downarrow},\label{hamil}
\eea
in which
\begin{itemize}
\item 
$t\in\RRR_+$ is the nearest neighbor hopping parameter, and is fixed to be $1$ in the rest of this paper.
$\Delta$ is the discrete Laplacian on the lattice: $\Delta {\bf a}^\pm_{\xx, \t}=\sum_{j=1,2}( {\bf a}^\pm_{\xx+\eee_j, \t}+{\bf a}^\pm_{\xx-\eee_j, \t}-2{\bf a}^\pm_{\xx, \t})$ and $\eee_j$, $j=1,2$, are the unit vectors in $\ZZZ^2$.

\item $\mu\in\RRR_+$ is the {\it renormalized} chemical potential. 

\item $\l\in\RRR$ is the strength of the on-site density-density interactions, called the {\it bare} coupling constant and $n_{\xx,\tau}={\bf a}^+_{\xx,\t}{\bf a}^-_{\xx, \t}$.
\end{itemize}

Let $\L_L^\star$ be the dual lattice of $\L_L$ with basis vectors $\bG_1=(2\pi,0)$, $\bG_2=(0,2\pi)$, the first Brillouin zone is defined as 
\be
\DD_L:=\RRR^2/\L^*_L=\big\{ \bk=\frac{n_1}{L}\bG_1+\frac{n_2}{L}\bG_2,\ {n_{1}, n_2\in[-\frac{L}{2}, \frac{L}{2}-1]\cap\ZZZ}\big\}.
\ee
The Fourier transform for the fermionic field operators are
$
{\bf a}^\pm_{\xx,\t}=\frac{1}{\b L^2}\sum_{\bk\in\DD_{L}}e^{\pm i\bk\cdot\xx}{\bf a}^\pm_{\bk,\t},$
and the inverse Fourier transform are given by ${\bf a}^\pm_{\bk,\t}=\sum_{\xx\in\L_L}e^{\mp i\bk\cdot\xx}{\bf a}^\pm_{\xx,\t}.$ The periodicity of ${\bf a}^\pm_{\xx,\t}$ on $\L_L$ implies that ${\bf a}^\pm_{\bk+n_1\bG_1+n_2\bG_2,\t}={\bf a}^\pm_{\bk,\t}$ and the commutation relations become:
\be
\{{\bf a}^+_{\bk,\t},{\bf a}^-_{\bk',\t'}\}=L^2\delta_{\bk,\bk'}\delta_{\tau,\tau'},\ \{{\bf a}^+_{\bk,\t},{\bf a}^+_{\bk',\t'}\}=0=\{{\bf a}^-_{\bk,\t},{\bf a}^-_{\bk',\t'}\}.\nn
\ee

Given the inverse temperature of the system $\b=1/T$, $T>0$, the Gibbs states associated with $H_{L}$ are defined by $\langle\cdot\rangle=\Tr_{\bF_L}\ [\ \cdot\ e^{-\beta H_{L}}]/Z_{\beta,\L_L}$, in which the normalization factor $Z_{\beta,\L_L}=\Tr_{\bF_L}e^{-\beta H_{L}}$ is called the partition function and the trace is taken w.r.t. the vectors in $\bF_L$. Define $\Lambda_{\beta, \Lambda_L}:=[-\b,\b)\times\L_L$. For $x_0\in [-\b,\b)$, the imaginary-time evolution of the fermionic operators are defined as ${\bf a}^\pm_{x}=e^{x_0H_L}{\bf a}^\pm_{\ \xx} e^{-x_0H_L}$, $x\in\L_{\b,L}$.

The $2p$-point Schwinger functions, $p\ge0$, are (formally) defined as:
\bea\label{nptsch}
&&S_{2p, \beta, L}(x_1,\e_1,\t_1;\cdots x_{2p} ,\e_{2p},\t_{2p};\lambda):=
\langle\bT\{ {\bf a}^{\e_1}_{x_1,\t_1}\cdots {\bf a}^{\e_{2p}}_{x_{2p},\t_{2p}}\}\rangle_{\beta,L}\nn\\
&&\quad\quad\quad:=\frac{1}{Z_{\beta,\Lambda_L}}\Tr_{\FFF_L}\Big[ e^{-\beta H_L}\bT\{{\bf a}^{\e_1}_{({x_1^0},\xx_1),\t_1}\cdots {\bf a}^{\e_{2p}}_{({x_{2p}^0},\xx_{2p}),\t_{2p}}\}\Big],
\eea
in which $\bT$ is the fermionic time-ordering operator, defined as
\bea
\bT\{ {\bf a}^{\e_1}_{(\xx_1, {x_1^0}),\t_1}\cdots {\bf a}^{\e_{2p}}_{(\xx_{2p}, {x_{2p}^0}),\t_{2p}}\}={\rm sgn} (\pi)\ {\bf a}^{\e_{\pi(1)}}_{(\xx_{\pi(1)}, {x}_{\pi(1)}^0),\t_{\pi(1)}}\cdots {\bf a}^{\e_{\pi({2p})}}_{(\xx_{\pi({2p})}, {x}_{\pi(n)}^0),\t_{\pi({2p})}},\nn
\eea
and $\pi$ is the permutation operator such that ${x}_{\pi(1)}^0\ge{x}_{\pi(2)}^0\ge\cdots\ge{x}_{\pi(2p)}^0$. If some fermionic operators are evaluated at equal time, the ambiguity is solved by putting ${\bf a}^-_{x_i,\t_i}$ on the right of ${\bf a}^+_{x_i,\t_i}$. We are particularly interested in the interacting two-point Schwinger function $S_{2,\b,L}(x,y;\l)$:
\bea
S_{2,\b,L}(x,y;\l)={\frac{1}{\b L^2}}
\sum_{k=(k_0, \kk)\in\DD_{\b, L}}e^{ik_0\cdot(x_0-y_0)+i\bk\cdot(\xx-\yy)} \hat S_{2,\beta}(k,\l)\label{fourier1},
\eea
in which the summation runs over the set $\DD_{\b, L}=\cD_\beta\times \cD_L$, in which  $\DD_{\b}:=\{\frac{2\pi}{\beta}(n+\frac12), n\in\NNN\}$ is the set of the components $k_0$, called the Matsubara frequencies. $\hat S_{2,\beta}(k,\l)$ is the two-point Schwinger function in the momentum space, {\it formally} defined as:
\bea\label{sfe1}
\hat S_{2,\beta}(k,\l)\delta_{\tau,\tau'}&:=&\langle{\bf T}\{{\bf a}^-_{k,\t}{\bf a}^+_{p,\t'}\}\rangle\\
&=&\delta_{\t,\t'}\delta(p-k)\frac{1}{ik_0-e_0(\bk,\mu)+\Sigma((k_0,\kk),\lambda)},\nn
\eea
and $\Sigma((k_0,\kk),\lambda)$ is called {\it the self-energy function}. The non-interacting  two-point Schwinger function (also called the free propagator) is defined as:
\bea
C_{\b}(x-y)&:=&\lim_{L\rightarrow\infty}S_{2,\b,L}(x,y;0)\nn\\
&=&\lim_{L\rightarrow\infty} {\frac{1}{\b L^2}}
\sum_{k=(k_0, \kk)\in\DD_{\b, L}}e^{ik_0\cdot(x_0-y_0)+i\bk\cdot(\xx-\yy)}\hat C(k_0,\bk)\label{free2pt},
\eea
in which \bea\label{2ptk}
\hat C(k_0,\bk)=\frac{1}{ik_0-e_0(\kk,\mu)},
\eea
is the free propagator in the momentum space, $e_0(\kk,\mu)=2-\cos k_1-\cos k_2-\mu$ is the band function. Let $\L^{\#}=\{\bb\in\RRR^2:\bb\cdot\xx\in2\pi\ZZZ\ {\rm for\ all}\ \xx\in\ZZZ^2\}=2\pi\ZZZ^2$ be the dual lattice of $\ZZZ^2$. In the limit $L\rightarrow\infty$, $\L_L\rightarrow\L=\ZZZ^2$, and $\cD_L\rightarrow\cB$, in which $\cB:=\RRR^2/\L^{\#}=(-\pi,\pi]^2$ is called the first Brillouin zone. So that the crystal momenta become continuous variables 
and the discrete sum ${\frac{1}{L^2}}\sum_{\kk\in\DD_{L}}$ becomes the integral $\int_{(-\pi,\pi]^2}d\bk$.
\begin{definition}\label{freefs}
When $k_0\rightarrow0$, the non-interacting Fermi surface defined by:
\be{\cal F}_0:=
\{\bk=(k_1, k_2)\in \cB\ \vert\ e_0(\bk,\mu)=0\},\label{freefs}\ee
\end{definition}
\begin{figure}[htp]
\centering
\includegraphics[width=0.32\textwidth]{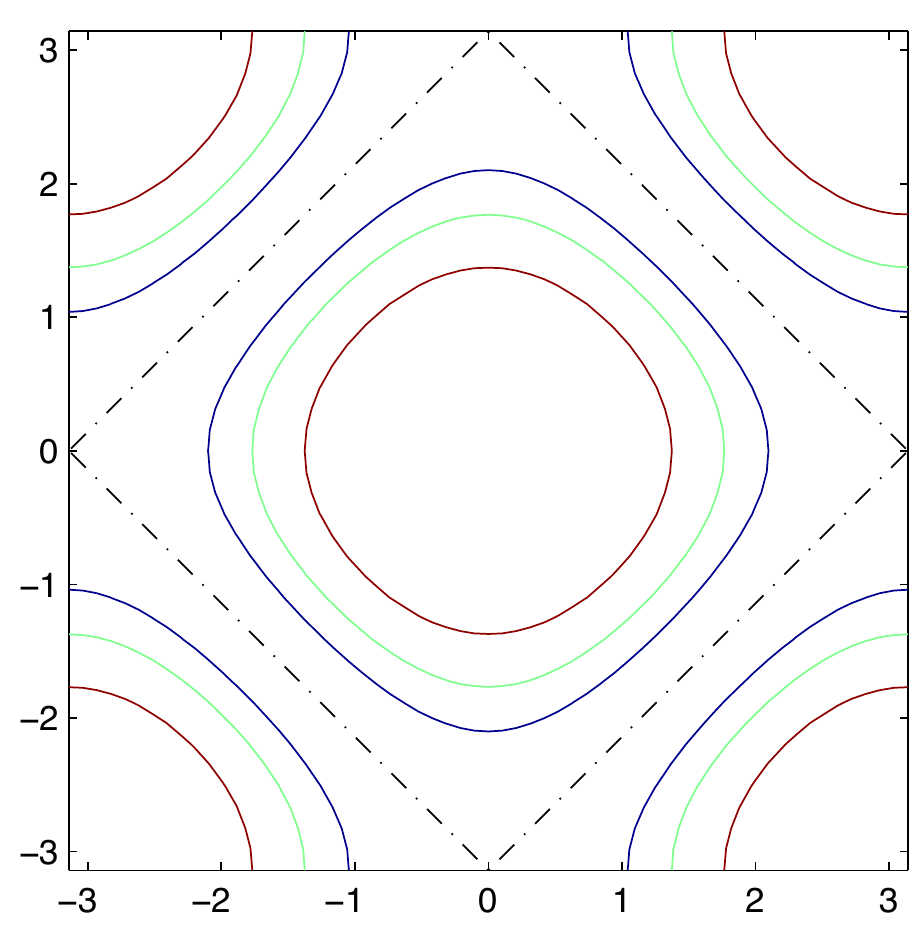}
\caption{\label{fs1} The non-interacting Fermi surfaces (thick lines) with $\mu=2-\mu_0$, for different values of $\mu_0$. The dotted square ($\mu_0=0$) is the Fermi surface at half-filling.
}
\end{figure}
\begin{definition}\label{intfs}
The interacting Fermi surface is defined by
\be\label{fint1}
\cF=\{\bk=(k_1, k_2)\in \cB\ \vert e(\bk,\mu,\l):=  e_0(\bk,\mu)-\Sigma(0,\bk,\l))=0\}.
\ee
\end{definition}
In this paper we consider the case in which the {\it renormalized} chemical potential $\mu=2-\mu_0$, in which $\mu_0=10^{-10}$ is fixed. The non-interacting Fermi surface is given by:
\be{\cal F}_0:=
\{\bk=(k_1, k_2)\in \cB\vert\ \cos k_1+\cos k_2-\mu_0=0\},\label{freefs}\ee
and the interacting Fermi surface is given by:
\be{\cal F}:=
\{\bk=(k_1, k_2)\in \cB\vert\ \cos k_1+\cos k_2-\mu_0+\Sigma(0,\bk,\l))=0\}.\label{ifs}\ee
Remark that, being a highly nontrivial function of the crystal momentum $\bk$, $\Sigma(0,\bk,\lambda)$ it is not known ahead of time and $\cF$ doesn't coincide with $\cF_0$ in general. This is the moving Fermi surface problem and would result in divergence in the perturbation expansion of the interacting Schwinger functions. We solve this problem by introducing the following counter-terms to the original many-body Hamiltonian so as to fix the interacting Fermi surface:
\bea
N_{L}(\lambda)&=&\frac{1}{L^2}\sum_{\bk\in\cD_L}\sum_{\t\in \{\uparrow,\downarrow\}}\Big[\ \delta\mu(\lambda)\ {\bf a}^+_{\bk,\t}{\bf a}_{\bk,\t}+
\hat\nu(\bk,\lambda){\bf a}^+_{\bk,\t}{\bf a}_{\bk,\t}\Big],
\eea
whose coefficients satisfy the following normalization conditions:
\be\label{rncd1a}
\delta\mu(0)=0,\quad \hat\nu(\bk,0)=0.
\ee
The (new) grand-canonical Hamiltonian is defined by:
\be
H_L=H_L^0+V_{L}+N_{L}.
\ee
Combining $H^0_L$ and $N_L$, we obtain the following quadratic term:
\bea
\sum_{\bk\in\cD_L}\sum_{\t\in \{\uparrow,\downarrow\}}{\bf a}^+_{\bk,\t}[ e_{bare}(\bk,\mu,\lambda)]{\bf a}_{\bk,\t},\nn
\eea
in which 
\be\label{bme}
e_{bare}({\bk,\mu,\l})=e_0(\bk,\mu)+\delta\mu(\lambda)+\hat\nu(\bk,\lambda)
\ee
is called the {\it bare band function} and $\mu_{bare}=\mu+\delta\mu(\lambda)$ is called the bare chemical potential. The Hamiltonian for the {\it new model} can be considered as
the one with band $e_{bare}$ and interaction potential $V_L(\lambda)$. Then the (new) interacting propagator is given by:
\bea\label{newintp}
\frac{1}{ik_0-e_0(\bk,\mu)+\delta \mu(\lambda)+\hat\nu(\kk,\l)+\Sigma(k,\lambda)},
\eea
and the singularities of the new interacting propagator \eqref{newintp} at $k_0=0$ are required to coincide with the non-interacting Fermi surface $\cF_0$, which put constraints on the counter-terms, also called {\it the renormalization conditions}. Notice that $\Sigma(k,\lambda)$ can be further
decomposed as: 
\be\label{selfdc}
\Sigma((k_0,\kk),\lambda)=T(\lambda)+\hat\Sigma((k_0,\kk),\lambda),
\ee 
in which $T(\lambda)$ is the localized part of $\Sigma((k_0,\kk,\l)$, called {\it the tadpole term}, is independent of the external momentum $\bk$ and $\hat\Sigma((k_0,\kk),\lambda)$ is the non-local part of $\Sigma((k_0,\kk,\l)$. 
By formally expanding \eqref{bme} as:
\be\label{intpr0}
\sum_{n=0}^\infty\frac{1}{ik_0-e_0(\bk,\mu)}\Bigg[\ \frac{\delta \mu(\lambda)+\hat\nu(\kk,\l)+\Sigma(k,\lambda)}{ik_0-e_0(\bk,\mu)}\ \Bigg]^n,
\ee
the renormalization conditions can be formulated as:
\begin{definition}[The renormalization conditions]\label{conj1}\\
\begin{itemize}
\item (a) The numerator in \eqref{intpr0} vanishes on the Fermi surface:
\bea
&&\delta\mu(\lambda)+T(\lambda)=0,\label{rncd1}\\ 
&&\hat\nu(P_F(\bk),\lambda)+\hat\Sigma((0,P_F(\bk)),\lambda )=0.\label{rncd2}
\eea
Here $P_F$ is a projection operator which maps each $\bk\in\cD_L$ to a unique $P_F\bk\in\cF_0$.
\item (b) the ratio
\be\label{rncd3}
\frac{\delta \mu(\lambda)+\hat\nu(\kk,\l)+\Sigma(k,\lambda)}{ik_0-e_0(\bk,\mu)}
\ee
is locally bounded for all $(k_0,\bk)\in\cD_{\b,L}$, up to a zero-measure set.
\end{itemize}
\end{definition}
We have the following theorem concerning the counter-terms (see also Theorem $7.4$):
\begin{theorem}\label{conj2}
There exist a pair of counter-terms $\delta \mu(\lambda), \hat\nu(\kk,\l)$, both are analytic functions of $\lambda$, such that the renormalization conditions introduced in Definition \ref{conj1} can be satisfied. In particular, $\hat\nu(\bk,\l)$ is $C^{1}$ but not $C^2$ differentiable w.r.t. the external momentum $\bk$.
\end{theorem}
\begin{remark}
Since the set of counter-terms $\{\delta \mu(\lambda), \hat\nu(\kk,\l)\}$ that satisfy conditions $(a)$ and $(b)$ maybe highly non-trivial, we have indeed defined a class of (bare) Hubbard models on the square lattice whose (renormalized) interacting Fermi surfaces coincide with $\cF_0$.
\end{remark}

In the following we will study the connected Schwinger functions $S^c_{2p,\beta}(\lambda)$, $p\ge0$, and the self-energy function $\Sigma(k, \lambda)$ in the {\it thermodynamic limit} $L\rightarrow\infty$, $\L_L\rightarrow\L=\ZZZ^2$. A fundamental mathematical problem concerns the well-definedness of these quantities and their regularities properties, or in other words, if the ground state of this model is a Fermi liquid.
\begin{definition}[Salmhofer's criterion of Fermi liquid]\label{salmc}
A $2$-dimensional many-fermion system at positive temperature is a Fermi liquid if the thermodynamic limit of the Schwinger functions in the momentum space, $\hat S_{2p}(k,\lambda)$, $p\ge1$, exist for $|\lambda|<\lambda_0(T)$ and if there are constants $C_0, C_1, C_2>0$ independent of $T$
and $\lambda$ such that the following holds. (a) The perturbation expansion for the momentum space self-energy $\hat\Sigma(k,\lambda)$ converges for all $(\lambda,T)$ with $|\lambda\log T|<C_0$. (b) The self-energy $\hat\Sigma(k, \lambda)$ satisfies the following regularity conditions:
\begin{itemize}
\item $\hat\Sigma(k, \lambda)$ is twice differentiable in $k_0$, $k_+,k_-$ and
\be
\max_{\beta=2}\Vert\partial_{k_{\a}}^\beta\hat\Sigma(k, \lambda)\Vert_{\infty}\le C_1,\ \a=0,\pm.
\ee
\item The restriction of the self-energy on the Fermi surface is $C^{\beta_0}$ differentiable w.r.t. the momentum, in which $\beta_0>2$, and
\be
\max_{\beta=\beta_0}\Vert\partial_{k_{\a}}^\beta\Sigma(k, \lambda)\Vert_{\infty}\le C_2,\ \a=0,\pm.
\ee
\end{itemize}
\end{definition} 

\subsection{The main result}
The main results are summarized in the following theorem (see also Theorem \ref{tpc}, Theorem \ref{cth1}, Theorem \ref{mqua}, Theorem \ref{maina} and Theorem \ref{mainb} for the precise presentation of the main results).
\begin{theorem}\label{mainthm}
Consider the square Hubbard model in which the renormalized chemical potential is fixed to be $\mu=2-\mu_0$, $\mu_0=10^{-10}$, at positive temperature $0<T\ll1$. There exist a pair of counter-terms $\delta \mu(\lambda), \hat\nu(\kk,\l)$ obeying the renormalization conditions in Definition \ref{conj1}, such that, after taking the thermodynamic limit $L\rightarrow\infty$, the perturbation series of the connected $2p$-point Schwinger functions, $p\ge1$, as well as the self-energy have positive radius of convergence in the domain $\RR_T:=\{\lambda\in\RRR\ |\ \vert\lambda\vert\le C_2/|\log (\mu_0T/C_1)|^2\}$, in which $C_1, C_2$ are positive constants independent of $T$ and $\lambda$. For any fixed $\l\in\RR_T$, the critical temperature $T_c$ for the phase transition is $T_c= \frac{C_1}{\m_0}\exp{(-C^{1/2}_2|\lambda|^{-1/2})}$. The second derivatives of the self-energy w.r.t. the external momentum are not uniformly bounded but are divergent for $T\rightarrow0$.
\end{theorem}

\begin{corollary}\label{phaset}
As a corollary of Theorem \ref{mainthm} (see also Theorem \ref{maina} and Theorem \ref{maine}), the perturbation series for the two-point Schwinger function is divergent if 
\be\mu_0T\le C_1e^{-(C_2/|\lambda|)^{1/2}},\ {\rm for\ any\ fixed}\ \l\in\RR_T,
\ee
which suggests that phase transition may happen with critical temperature 
\be T_c:=\begin{cases}\frac{C_1}{\mu_0}e^{-(C_2/|\lambda|)^{1/2}},\ {\rm for}\ \m_0\ge T\ {\rm fixed},\\
C_1'e^{-(C'_2/|\lambda|)^{1/2}},\ {\rm for}\ \m_0\le T,\end{cases}\label{crit}
\ee
in which $C'_1=C_1^{1/2}$ and $C'_2=C_2/4$.
\end{corollary}

\begin{remark}\label{cbgm}
Remark that the techniques employed in this paper is not suitable for $\mu_0$ large, say, $\mu_0\in (1.7,2)$, and our result can't be reduced to the one in \cite{BGM2}, in which the authors proved that the ground state of the square Hubbard with $\mu\le0.3$ is a Fermi liquid for $T\ge T^1_c=\tilde C_1e^{-\tilde C_2/|\l|}$, with $\tilde C_1$ and $\tilde C_2$ two positive constants independent of $T$ and $\l$. The reason is the following. While the Fermi surface considered in \cite{BGM2} is close to a circle so that both the isotropic sectors and the non isotropic sectors (\cite{BGM2}, Section 2.5) can be used to pave the shells surrounding the Fermi surface, the Fermi surface considered in the present paper is close to a square such that the curvature radius may change drastically at different points on the Fermi surface. So we have to introduce the highly non isotropic sectors (see Definition \ref{defsec}) to obtain the desired decay for the propagator, which leads to a different sector analysis. 
\end{remark}

\section{The Multi-scale Analysis}
\subsection{The Fermionic functional integrals}
The first step in the renormalization group analysis is to express the Schwinger functions in terms of fermionic functional integrals, also called the Berezin integrals, which are linear functionals on the Grassmann algebra ${\bf Gra}$ generated by the Grassmann variables $\{\hat\psi^\e_{k,\t}\}^{\t=\uparrow\downarrow;\epsilon=\pm}_{k\in\DD_{\b,L}}$, 
which satisfy the periodic condition in the momentum variables: $\hat\psi^\e_{k_0,\bk+n_1{\bf G}_1+n_2{\bf G}_2,\t}=\hat\psi^\e_{k_0,\bk,\t}$, but anti-periodic condition in the frequency variable: $\hat\psi^\e_{k_0+\beta,\bk,\t}=-\hat\psi^\e_{k_0,\bk,\t}$.
The product in ${\bf Gra}$ is defined by: $\hat\psi^\e_{k,\t}\hat\psi^{\e'}_{k',\t'}=-\hat\psi^{\e'}_{k',\t'}\hat\psi^\e_{k,\t}$, for  $(\e,\t,k)\neq(\e',\t',k')$ and $(\hat\psi^\e_{k,\t})^2=0$. Define a measure $D\psi=\prod_{k\in\DD_{\b, L}, \\
\t=\pm}d\hat\psi_{k,\t}^+
d\hat\psi_{k,\t}^-$ on ${\bf Gra}$, also called the Grassmann Lebesgue measure and let $Q( \hat\psi^-, \hat\psi^+)$ be a monomial function of $\hat\psi_{k,\t}^-, \hat\psi_{k,\t}^+$, the Grassmann integral $\int Q D\psi$ is defined to be $1$ for $Q( \hat\psi^-, \hat\psi^+)=\prod_{k\in\DD_{\b, L}, \\
\t=\pm} \hat\psi^-_{k,\t} \hat\psi^+_{k,\t}$, up to a permutation of the variables, and $0$ otherwise. The Grassmann differentiation is defined by 
${\partial_{ \hat\psi^\e_{k,\t}}}{ \hat\psi^{\e'}_{k',t'}}=\delta_{k,k'}\delta_{\t,\t'}\delta_{\e,\e'}$, which also satisfy the anti-commutation relation $\partial_{ \hat\psi^\e_{k,\t}}\partial_{ \hat\psi^{\e'}_{k',\t'}}=-\partial_{ \hat\psi^{\e'}_{k',\t'}}\partial_{ \hat\psi^\e_{k,\t}}$ for $(\e,k,\t)\neq(\e',k',\t')$, and equals to zero otherwise.
The {\it Grassmann Gaussian measures} $P(d\psi)$ with covariance $\hat C(k)$ is defined as:
\be
P(d\psi) = (\NN)^{-1} D\psi \cdot\;\exp \Bigg\{-\frac{1}{\b L^2} \sum_{k\in\DD_{\b, L},\t={\uparrow\downarrow} } 
\hat\psi^{+}_{k,\t}{\hat C({k})}^{-1}\hat\psi^{-}_{k,\t}\Bigg\}\;,
\label{ggauss}\ee
where
\be
\NN=\prod_{\kk\in\DD_L,\t={\uparrow\downarrow}}{\frac{1}{\b L^2}}
[i k_0-e_0(\bk,\mu_0)] \label{norma}
\ee
is the normalization factor. The lattice Grassmann fields are defined as:
\be
\psi^\pm_{x,\t}=\sum_{k\in\DD_{\b, L}}
e^{\pm ikx}\hat\psi^\pm_{k,\t},\ \ x\in\Lambda_{\beta,L}.\nn
\ee

The interaction potential becomes:
\bea 
\VV_L(\psi,\lambda)=
\l\ \int_{\Lambda_{\beta,L}} d^3x \ \psi^+_{x,\uparrow}
\psi^-_{x,\uparrow}\psi^+_{x,\downarrow}
\psi^-_{x,\downarrow}\label{potx}+\frac{1}{\b L^2}\sum_{k\in\DD_{\b, L}}\sum_{\tau=\uparrow, \downarrow}[\delta\mu(\lambda)+\nu(\kk,\lambda)]
\hat\psi^+_{k,\t}\hat\psi^-_{k,\t},\nn
\eea
where $\int_{\Lambda_{\beta,L}} d^3x:=\int_{-\beta}^\beta dx_0\ \sum_{\xx\in\L_L}$ is a short-handed notion for the integration and sum. Define the non-Gaussian measure $P^I_L(d\psi):=P(d\psi)e^{-\VV_L(\psi)}$ over ${\bf Gra}$ and let $P^I(d\psi)=\lim_{L\rightarrow\infty}P^I_L(d\psi)$ be the limit of the sequence of measures indexed by $L$ (in the topology of weak convergence of measures), whose total mass $Z:=\int P^I(d\psi)$ is called the partition function. Let $\{\eta_{\xi}^+,\eta_{\xi}^-\}$ be another set of Grassmann fields, called the external fields, in which $\xi_i=(x_i,\tau_i)$, we can define the moment generating functional by
\be
Z(\{\eta^+,\eta^-\})=\frac{1}{Z}\int e^{-\VV(\psi,\lambda)+\sum_{\xi_i}(\eta^+_{\xi_i}\psi^-_{\xi_i}+\psi^+_{\xi_i}\eta^-_{\xi_i})} P(d\psi).
\ee
The $2p$-point Schwinger functions are defined as the moments of the measure $P^I(d\psi)$:
\bea
&&S_{2p,\b}(\xi_1,\cdots,\xi_{2p};\l)=\frac{\partial^{2p}}{\prod_{i=1}^p\partial \eta^+_{\xi_i}\partial \eta^-_{\xi_i}\ }Z(\{\eta^+,\eta^-\})\vert_{\{\eta^\pm_{\xi_i}\}=0}\\
&=&\lim_{L\rightarrow\infty}\frac{\int\psi^{\epsilon_1}_{x_1,\t_1}\cdots \psi^{\epsilon_{2p}}_{x_{2p},\t_{2p}} P(d\psi)e^{-\VV_L(\psi,\lambda)}}{\int P(d\psi)e^{-\VV_L(\psi,\lambda)}}=:\frac{\int\psi^{\epsilon_1}_{x_1,\t_1}\cdots \psi^{\epsilon_{2p}}_{x_{2p},\t_{2p}} P(d\psi)e^{-\VV(\psi,\lambda)}}{\int P(d\psi)e^{-\VV(\psi,\lambda)}}.\nn
\eea
We assume in the rest of this paper that the thermodynamic limit $\L_L\rightarrow\L=\ZZZ^2$ has been already taken and will drop the parameter $L$. This assumption is justified by the rigorous construction of the connected Schwinger functions in the limit $L\rightarrow\infty$. The generating functional for the {\it connected} Schwinger functions is defined as
\be
W(\eta^+,\eta^-)=\log\frac{1}{Z}\int e^{-\VV(\psi,\lambda)+\eta^+\psi^-+\psi^+\eta^-} P(d\psi),
\ee 
the normalization factor $Z=\int e^{-\VV(\psi,\lambda)} P(d\psi)$ is also called the partition function. The {\it connected} $2p$-point Schwinger functions, which are the {\it commulants} of the of the probability measure $\frac{1}{Z}e^{-\VV(\psi,\lambda)+\eta^+\psi^-+\psi^+\eta^-} P(d\psi)$,  
are (formally) defined by:
\be
S^c_{2p}(\xi_1,\cdots,\xi_{2p})=\prod_{i=1}^p\frac{\delta^2}{\delta\eta^-(\xi_i)\delta\eta^+(\xi_i)}W(\{\eta^+,\eta^-\}).
\ee
By definition, the generating functional $W$ exists when the norm 
\be
\Vert S^c_{2p}\Vert_{L^\infty}=\max_j\sum_{\xi_j}\int_{i\neq j}d\xi_i\vert  S^c_{2p}(\xi_1,\cdots,\xi_{2p})\vert
\ee
of each connected Schwinger function $S^c_{2p}$, $p\ge1$, is finite. Define also the Grassmann fields $\{A^\epsilon,A_0^\epsilon\}$, $\epsilon=\pm$, which are conjugate to $\eta^+,\eta^-$, by:
\be\label{leg1}
A^\epsilon(\xi_i)+A_0^\epsilon(\xi_i)=\frac{\delta}{\delta \eta^\epsilon(\xi_i)}W(\eta^+,\eta^-),
\ee
where $A_0^\epsilon(\xi_i)$ is independent of $ \eta^\epsilon$ and is chosen in such a way that $A^\epsilon=0$ when $\eta^\epsilon=0$. The generating functional for the one-particle irreducible (1PI) functions, $\Gamma(A^+, A^-)$, is defined as the Legendre transform of $W(\eta^+,\eta^-)$ (see, eg. \cite{Feld1}):
\be
\Gamma(A^+,A^-)=W(\eta^+,\eta^-)-\sum_{\e=\pm}\frac{\delta}{\delta \eta^\epsilon}W(\eta^+,\eta^-)\eta^\e\vert_{\eta^\e=\eta^\e(A^+,A^-)},
\ee
where $\eta^e(A^+,A^-)$ refers to the inverse map $\eta^\e\rightarrow A^\e$ defined in \eqref{leg1}. The self-energy is (formally) defined as:
\be\label{self1}
\Sigma(\xi_1,\xi_2)=\frac{\delta^2}{\delta {A^+(\xi_1)}\delta {A^-(\xi_2)}}\ \Gamma(A^+,A^-)\vert_{A^\pm=0}.
\ee

\subsection{Scale Analysis}
The lattice structure plays the role of the short-distance cutoff for the spatial momentum, so that the ultraviolet behaviors of the Schwinger functions are rather trivial. The two-point Schwinger function is not divergent but has a discontinuity at $x_0=0$, $\xx=0$. Although summation over all scales of the tadpole terms is not absolutely convergent for $k_0\rightarrow\infty$, this sum can be controlled by using the explicit expression of the single scale propagator \cite{BG}. So we omit the ultraviolet analysis but introduce a suitable ultraviolet (UV) cutoff function $U(\bk)$, $\bk\in \RRR^2$, which is smooth and compactly supported. This can keep the momentum $\bk$ bounded. We consider only the infrared behaviors, which correspond to the cases of $T\ll1$ and the momenta getting close to the Fermi surface. It is mostly convenient to choose the infrared cutoff functions as the Gevrey class functions.
\begin{definition}\label{gev}
Given $\DD\subset\RRR^d$ and $h>1$, the Gevrey class $G^h_0(\DD)$ of functions of index $h$ is defined as the set of smooth functions $\phi\in\cC^\infty(\DD)$ such that for every compact subset $\DD^c\subset\DD$, there exist two positive constants $A$ and $\g$, both depend on $\phi$ and the compact set $\DD^c$, satisfying:
\bea
\max_{x\in K}|\partial^\alpha \phi(x)|\le A\g^{-|\alpha|}(|\alpha!|)^h,\ \alpha\in\ZZZ^d_+,\ |\alpha|=\alpha_1+\cdots+\alpha_d.\nn
\eea
The Gevrey class of functions with compact support is defined as: 
$G_0^h(\DD)=G^h(\DD)\cap C^\infty_0(\DD)$.
The Fourier transform of any $\phi\in G_0^h$ satisfies
\be
\max_{k\in\RRR^d}|\hat \phi(k)|\le Ae^{-h(\frac{\g}{\sqrt{d}}|k|)^{1/h}}.\nn
\ee
\end{definition}

The infrared cutoff function $\chi\in G^h_0(\RRR)$ is defined by:
\be
\chi(t)=\chi(-t)=
\begin{cases}
=0\ ,&\quad {\rm for}\quad  |t|>2,\\
\in(0,1)\ ,&\quad {\rm for}\quad  1<|t|\le2,\\
=1,\ &\quad {\rm for}\quad  |t|\le 1. 
\end{cases}\label{support}
\ee
Given any fixed constant $\gamma\ge10$, define the following partition of unity:
\bea\label{part1}
1&=&\sum_{j=0}^{\infty}\chi_j(t),\ \ \forall t\neq 0;\\
\chi_0(t)&=&1-\chi(t),\ 
\chi_j(t)=\chi(\gamma^{2j-2}t)-\chi(\gamma^{2j}t),\ {\rm for}\ j\ge1.\nn
\eea
The support $\cD_j$ of the cutoff function $\chi_j$ for $j\ge1$ is a shell surrounding 
$\cF_0$:
\be\label{multi1}
\cD_j=\Big\{k=(k_0,\kk)\ \vert\ \g^{-2j-2}\le k_0^2+e^2(\kk,1)\le 2\g^{-2j}\Big\}.\ee
 
Define also 
$$\chi^{(\le j)}(t)=\sum_{i=0}^j\chi_{i}(t),\ \chi^{(> j)}(t)=\sum_{i=j+1}^\infty\chi_{i}(t).$$


\begin{definition}
The free propagators in the infrared region are defined as:
\bea\label{irprop}
\hat C^{ir}(k)&=&\hat C(k)[U(\bk)-\chi^{(\le 0)}(k_0^2+e_0^2(\kk,\mu_0))],\nn\\
&=&\sum_{j=0}^\infty \ \hat C_j(k)\\
\hat C_j(k)&=&\hat C(k)\cdot \chi_j[k_0^2+e_0^2(\kk,\mu_0)].\label{irprop1}
\eea
\end{definition}
\begin{remark}
Remark that, it is always possible to choose a proper $U(\bk)$ such that $U(\bk)\ge \chi^{(\ge j)}[k_0^2+e_0^2(\kk,\mu_0)],$ for all $j\ge0$ and all $k=(k_0,\bk)\in\RRR\times \RRR^2$. 
\end{remark}

\begin{definition}\label{wholeprop}
Define the infrared cutoff index $j_{max}=\EEE(\tilde j_{max})$, in which $\tilde j_{max}$ is the solution to the equation $\gamma^{\tilde j_{max}-1}= 1/\sqrt2\pi T$ and
$\EEE(\tilde j_{max})$ is the integer part of $\tilde j_{max}$. Then the infrared propagator with cutoff index $j_{max}$ is defined as:
\bea\label{irpropm}
\hat C^{ir,j_{max}}(k)=\hat C(k)[U(\bk)-\chi^{(\le 0)}(k_0^2+e_0^2(\kk,\mu_0))-\chi^{(> j_{max})}(k_0^2+e_0^2(\kk,\mu_0))].
\eea
\end{definition}

It is convenient to study the geometry of the F.S. in a new coordinate basis $(e_+,e_-)$, $e_+=1/2(1,1)$, $e_-=1/2(-1,1)$, which is orthogonal but not normal. If we call $\tilde\bk=(k_+,k_-)$ the coordinates in the new basis, the Fermi surface is given by
\be
\cF_0=\{(k_+,k_-)\in\RRR^2,2\cos \frac{k_+}{2} \cos \frac{k_-}{2}-\mu_0=0\}.
\ee
\begin{notation}
We will drop the tilde in the variables $\tilde\bk$ can call them $\bk=(k_+,k_-)$ if no confusion can be made.
\end{notation}
\begin{figure}[htp]
\centering
\includegraphics[width=0.46\textwidth]{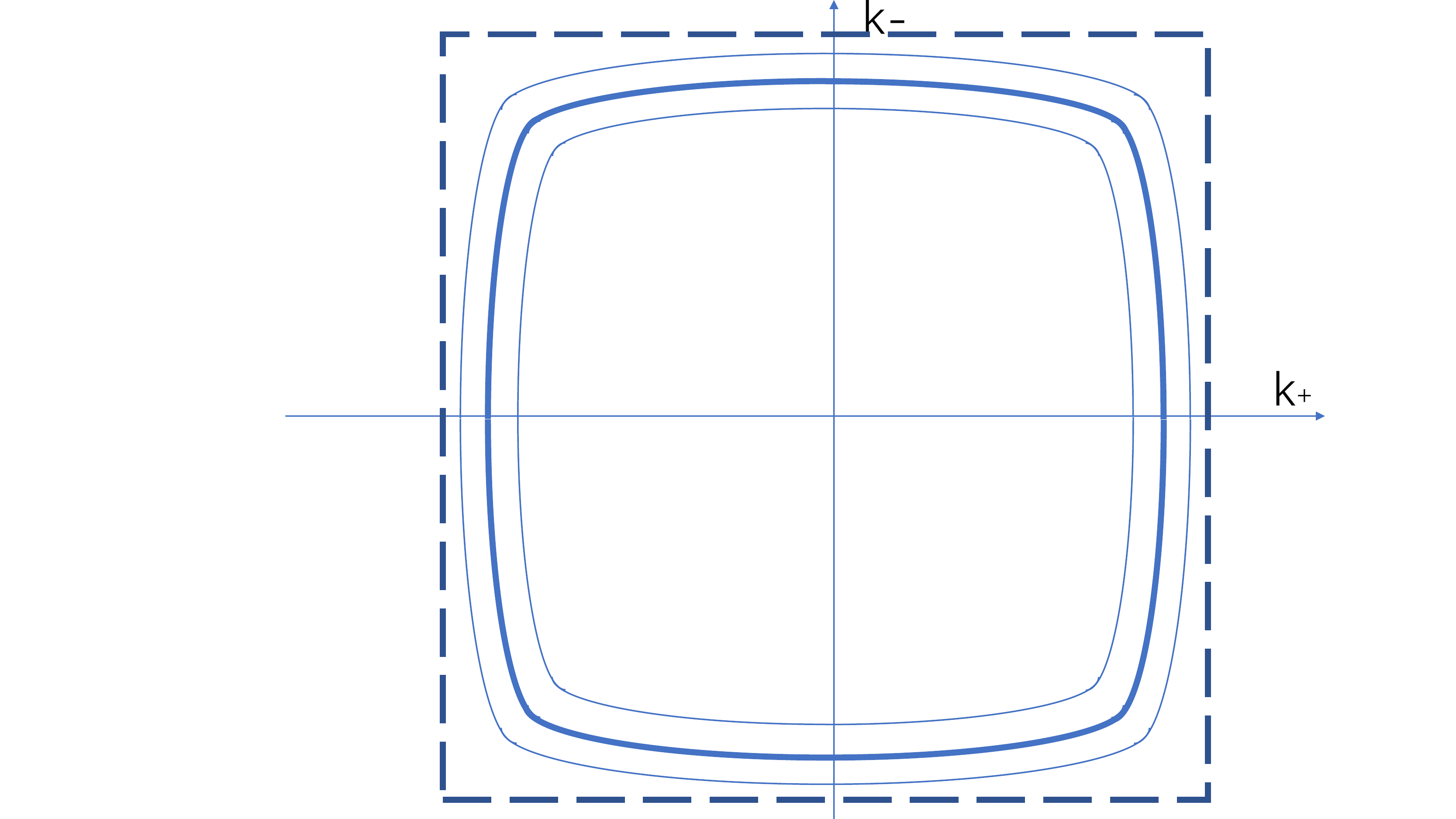}
\caption{\label{fs1} The Fermi surface (thick line) with $\mu=2-\mu_0$ and a shell (dotted line). The square is the Fermi surface with $\mu_0=0$ corresponding to the half-filling.
}
\end{figure}
The geometrical properties of a shell $\cD_j$ (cf. \eqref{multi1}) for fixed $\mu_0$ and $j\ge0$ has been discussed in detail in the Appendix. In each shell $\cD_j$, the term $k_0^2+e_0^2((k_+,k_-),\mu_0)$ is bounded by
$O(1)\gamma^{-2j}$, but the term $e_0^2((k_+,k_-),\mu_0)$, which can be of order $\gamma^{-2i}$ with $i\ge j$, is not fixed. Define $\bq=(q_+,q_-)$ by $\bk=\bk^F+\bq$, in which
$\bk^F=P_F(\bk)\in\cF_0$ is the unique projection of $\bk$ onto $\cF_0$. In order to obtain the optimal decaying bounds for the direct-space propagators, we need to control the size of $e_0^2((k^F_++q_+,k_-^F+q_-),\mu_0)$, by slicing the momentum space at each scale $j$ into smaller regions, called the {\it sectors}. Since $R_{max}\gg R_{min}$ for $\mu_0\ll1$, and $j\gg1$, by Proposition \ref{geom}, it is not convenient to employ the sectors defined in \cite{FMRT} or \cite{BGM2}, but we shall introduce the highly non isotropic sectors (see also \cite{RWang1}), as follows.
\begin{definition}[Sectors]\label{defsec}
For any fixed $\mu_0$ such that $0<\mu_0\ll1$, define $j_0=\EEE(\vert\log_\gamma \mu_0\vert)$. Given a set of integers $s\in[0, j]\cap\NNN$, define the following functions of partition of unity:
\bea\label{secf}
1=\sum_{s=0}^{j}v_{s}(r),\ \ \begin{cases}
v_{s}(r)=\chi_{s+1}(r),\ {\rm for}\ \ 0\le s\le j-1,\\
v_j(r)=\chi(\g^{2j}r).
\end{cases}
\eea
The free propagator at any slice $j$ can be decomposed as:
\bea
&&\hat C_j(k_0,k_+,k_-)=\sum_{\s=(s_+, s_-)}\hat C_{j,\s}(k_0,k_+,k_-)\label{sec0},\\
&&\hat C_{j,\s}(k)=\hat C_j(k)\cdot v_{s_+}(\vert q_+\vert^2)\ v_{s_-}(\vert q_-\vert^2),\label{sec1a}
\eea
in which $s_+\in[\max(0,\frac j2-\frac{j_0}{2}), j]\cap\NNN$, $s_-\in[\max(0,\frac j2-\frac{j_0}{2}), j]\cap\NNN$, are two integers satisfying the following constraint:
\be s_++s_-\ge\frac32 j-\frac{j_0}{2}\label{constr1}.
\ee
 $\sigma=(s_+,s_-)$ is called a sector index, and $\hat C_{j,\s}(k)$
is called a sectorized propagator. The support of a sectored propagator is a set of rectangles whose sides are parallel to the axis $k_+=0$ and $k_-=0$. Each such rectangle $\Delta^j_{s_+,s_-}$ is called a sector with scale index $j$ and sector index $\sigma=(s_+,s_-)$. Correspondingly, we have the following decomposition of the fermionic fields
\be\label{decfield}
\hat\psi^\pm(k_0,\bk^F+\bq)=\sum_j\sum_{\s=(s_+,s_-)}\hat\psi^{\pm(j,\s)} (k_0,\bk^F+\bq).
\ee

\end{definition}
\subsection{Constraints on the Sector Indices}
In this part we consider the possible constraints on the sector indices placed by conservation of momentum. By Definition \ref{defsec} we have $\frac{3j}{2} -\frac{j_0}{2}\le s_++s_-\le 2j$.
Let $\bk$ be a vector in the $j$-th neighborhood of $\cF_0$. Let $\s=(s_+,s_-)$ be the sector indices of $\bk$. In order that a sliced propagator $\hat C_{j,\s}(k_0,\bk)\neq0$, the momentum $k=(k_0,\bk^F+\bq)$ must satisfy the following bounds:
\be
\frac{1}{4}\gamma^{-2j-2}\le k^2_0\le \frac12 \g^{-2j},\nn
\ee
and
\be
\g^{-s_+-1}\le\vert q_\pm\vert\le\sqrt2\g^{-s_+},\quad \  s_\pm\in[\frac j2-\frac{j_0}{2}, j].\nn
\ee
\vskip.5cm

\begin{lemma}\label{secmain}
Let $\bk^{i}\in \RRR^2$ be the momentum entering or existing a vertex such that 
\be
\vert \bk^{1}+\cdots+\bk^{4}\vert \le c\g^{-j},\nn
\ee 
in which $j\ge0$, $c>0$ and $\g\ge10$ are two fixed constants. Let $\bk_F^{i}=P_F(\bk^{i})\in\cF_0$ be the unique projection on the Fermi surface and $\bq^{i}=(q^{i}_{+}, q^{i}_{-})$ be the quasi-momentum. Let $\Delta^j_{s_+,s_-}$, $i=1,\cdots,4$, be the four sectors of length $\g^{-s^{i}_+}$ and height $\g^{-s^{i}_-}$ such that 
$\bq^{i}\in \Delta^j_{s_+,s_-}$. Suppose that, among all the sector indices in $+$ direction, $s_{+}^{1}, s_{+}^{2}$ are the two smallest sector indices and $s_{+}^{1}\le s_{+}^{2}$. Then either  $|s_{+}^{1}-s_{+}^{2}|\le1$ or $s_{+}^{1}=j_1$, in which $j_1$ is strictly smaller then $j_2, j_3$ and $j_4$. We have exactly the same results for the sector indices $s_{-}^{i}$, $i=1,\cdots,4$.
\end{lemma}
\begin{proof}
First of all we consider the possible constraints for the quasi momentum $q^{i}_{+}$ and $q^{i}_{-}$, $i=1,\cdots,4$. Since $\bq^{i}\in\Delta^j_{s_+,s_-}$, we can always arrange sector indices in the order $s_{1,+}\le s_{2,+}\le s_{3,+}\le s_{4,+}$ and the scale indices in the order
$j_1\le j_2\le j_3\le j_4$. Then either $s_{1,+}<j_1$ or $s_{1,+}=j_1$. In both cases we have (cf. Formula \eqref{sec1a}) $|q_{i,+}|\le\sqrt2\g^{-s_{2,+}}$, for $i=2,3,4$, and $|q_{1,+}|\ge{\g^{-s_{1,+}-1}}$. In order that the equation $q_{1,+}+q_{2,+}+q_{3,+}+q_{4,+}=0$ holds, we must have
\be
2{\g^{-s_{1,+}-1}}/{\pi }\le \frac{6\sqrt2}{\pi}\g^{-s_{2,_+}},\nn
\ee
which implies
\be
s_{2,+}\le s_{1,+}+1+\log_\g (3\sqrt2).\nn
\ee
For any $\g\ge 10$, we have $0<\log_\g (3\sqrt2)<1$. Since $s_{1,+}$ and $s_{2,+}$ are integers, we have $|s_{2,+}-s_{1,+}|\le1$. Following the same arguments we can prove the same result for sectors in the $"-"$ direction.
\end{proof}
%
\subsection{Decay properties of the sectorized propagators in the direct space}
In this section we study the decaying behaviors of the free propagators in the direct space.
First of all, define the dual coordinates to the momentum $k_+$, $k_-$, as: 
\be
x_+=(x_1+x_2)/2,\ x_-=(-x_1+x_2)/2.\nn
\ee
Let $(j, \sigma)=(j,s_+,s_-)$ be the scale index and sector indices for a sector. It is useful to introduce a new index, $l(j,\s)$, called the $depth$ of a sector, which describes the distance of a sector with index $s$ to the Fermi surface, by \be l(j,\s)=s_++s_--3j/2+j_0/2.\label{lindex}\ee
\begin{lemma}\label{bdx1}
Let $C_{j,\sigma}(x-y)$ be the Fourier transform of $\hat C_{j,\sigma}(k_0,k_\pm)=C_{j,\sigma}(k_0,q_++k_{F,+},q_-+k_{F,-})$, (c.f. Eq. \eqref{sec1a}), $x=(x_0,x_+,x_-)$. There exists a constant $K$, which depends on the model but independent of the scale indices and the sector indices, such that:
\be\label{decay1}
\Vert C_{j,\sigma}(x-y)\Vert_{L^\infty}\le K \g^{-l(j,\s)-3j/2+j_0/2}\ e^{-c[d_{j,\s}(x,y)]^\a},
\ee
where 
\be\label{dist0}
d_{j,\s}(x,y)=\g^{-j}\vert x_0-y_0\vert+\g^{-s_+}\vert x_+-y_+\vert+\g^{-s_-}\vert x_--y_-\vert,
\ee
$\a=1/h\in(0,1)$ is the index characterizing the Gevrey class of functions (cf. Definition \ref{gev}).
\end{lemma}
\begin{proof}
First of all, 
\be\label{tad1}
\Vert\hat C_{j,\s}(k)\Vert_{L^\infty}:=\sup_{k\in \cD_j}\vert\hat C_{j,\s}(k)\vert\le K\g^{j},
\ee
in which $K$ is a positive constant independent of the scale index, and the integration measure $\int dk_0dq_-dq_+$ constrained to a sector is bounded by $\g^{-j}\cdot \g^{-s_+}\cdot\g^{-s_-}$. So we obtain the pre-factor of \eqref{decay1}. Let $\frac{\partial}{\partial k_0}f=(1/2\pi T)[f(k_0+2\pi T)-f(k_0)]$ be the difference operator.  To prove the decaying behavior of \eqref{decay1}, it is enough to prove that
\bea\label{decay4}
\Vert\frac{\partial^{n_0}}{\partial k_0^{n_0}}\frac{\partial^{n_+}}{\partial (q_+)^{n_+}}\frac{\partial^{n_-}}{\partial (q_-)^{n_-}} C_{j,\s}(k_0, q_+,q_-)\Vert_{L^\infty}
\le K^n\g^{jn_0}\g^{s_+n_+}\g^{s_-n_-}(n!)^{1/\alpha},
\eea
where $n=n_0+n_++n_-$ and $\Vert\cdot\Vert$ is the sup norm. We can easily prove that there exist some positive constants $K_1, K_2, K_3$ such that $\Vert\frac{\partial}{\partial q_-} v_{s_-}(q^2_-)\Vert\le K_1\g^{s_-}$; when the operator $\frac{\partial}{\partial q_-}$ acts on $\chi_j[k_0^2+e^2(q_+,q_-)]$, the resulting term is simply bounded by $K_2\g^{-s_-}$; when $\frac{\partial}{\partial q_-}$ acts on $[-ik_0+e(q_+,q_-)]^{-1}$, the resulting term is bounded by $K_3\g^{-s_-}$. Similarly, each operator $\frac{\partial}{\partial q_+}$ acts on various terms of $C_{j,\s}(k_0,q_+,q_-)$ is bounded by $K\g^{s_+}$. Finally, each derivation $\frac{\partial}{\partial k_0}$ on the propagator gives a factor $\g^j$.
The factor $(n!)^{1/\alpha}$ comes from derivations on the compact support functions, which are Gevrey functions of order $\alpha$. When $j=j_{max}$, the propagator decays only in the $x_0$ direction but not in the $x_+$ or $x_-$ direction. Let $K=K_1\cdot K_2\cdot K_3$, the result of this Lemma follows.
\end{proof}
\begin{lemma}
There exists some positive constant $K$ which is independent of $j$, $\s$ such that
\be\label{tad2}
\Big\Vert\ C_{j,\sigma}(x)\ \Big\Vert_{L^1}\le K\g^{j}.
\ee
\end{lemma}
\begin{proof}
This lemma can be proved straightforwardly using Lemma \ref{bdx1}. We have
\bea
&&\Big\Vert C_{j,\sigma}(x)\ \Big\Vert_{L^1}\quad\le K\ \Big|\ \int_{\Lambda_{\beta}} dx_0 dx_+ dx_- \  C_{j,\sigma}(x)\ \Big|\nn\\
&&\quad\le K\g^{-j-l}\g^{(j+s_++s_-)}\le K\g^{j}.
\eea
\end{proof}
Remark that, comparing to the $L^\infty$ norm for a sliced propagator $C_{j,\sigma}(x)$, 
a factor $\g^{j+s_++s_-}$ is lost when taking the $L^1$ norm. It is convenient to define a new scale index.
\begin{definition}\label{indexr}
Define the scale index $r=\EEE(\frac{j+s_++s_-}{2})$. We have $r\ge r_{min}:=5j_{min}/4-j_0/4\ge0$ and $r_{max}(T):=\EEE(\frac{3}{2}\ j_{max}(T))$.
Correspondingly, we have the following decomposition for the propagator:
\be
\hat C(k_0,q_+,q_-)=\sum_{r=0}^{r_{max}(T)}\sum_\s\hat C_{r, \s}(k_0,q_+,q_-).
\ee
In terms of the scale index $r$, a sector is also denoted by $\Delta^r_{s_+,s_-}$. Correspondingly, we have the following decomposition for the fermionic fields:
\be\label{decfield2}
\hat\psi^\pm(k_0,\bq+\bk^F)=\sum_r\sum_{\s=(s_+,s_-)}\hat\psi^{\pm(r,\s)} (k_0,\bq+\bk^F).
\ee
\end{definition}
Since $|x-\EEE(x)|\le1$, $\forall x\in\RRR$, we shall simply forget the integer part $\EEE(\cdot)$ in the future sections. The four indices $j$, $s_+$, $s_-$ and $r$ are related by the relation $r=\frac{j+s_++s_-}{2}.$
With the new index $r$, the constraints for the sector indices becomes $s_++s_-\ge 6r/5-j_0/5$ and $0\le r\le r_{max}=3j_{max}/2$.

\section{The Perturbation Expansion}
\subsection{The BKAR jungle formula and the power-counting theorem}
In this section we study the perturbation expansions for the Schwinger functions. Recall that a general $2p$-point Schwinger function at temperature $\b=1/T$ is defined as:
\bea
&&S_{2p,\b}(\lambda; x_1,\t_1;\cdots,x_{2p},\t_{2p},\l)\\
&&=\frac{1}{Z}\int d\mu_C(\bar\psi,\psi)\Big[ \prod_{i=1}^p\prod_{\e_i=\pm,\t_i}\psi^{\e_i}_{\tau_i}(x_i)\ \Big]\ \Big[
\prod_{i=p+1}^{2p}\prod_{\e_i=\pm,\t_i}\psi^{\e_i}_{\tau_i}(x_i)\Big]e^{- {\VV}(\bar\psi,\psi)}.\nn
\eea
Let $\{\xi^x_i=(x_{i},\t_{i})\}$, $\{\xi^y_i=(y_{i},\t_{i})\}$ and $\{\xi^z_i=(z_{i},\t_{i})\}$ be set of indices associated with the Grassmann variables $\{\psi^{\e_i}_{\tau_i}(x_i)\}$, $\{\psi^{\e_i}_{\tau_i}(y_i)\}$ and $\{\psi^{\e_i}_{\tau_i}(z_i)\}$, respectively. Expanding the exponential into power series and performing the Grassmann integrals, we have
\bea\label{part2p}
&&S_{2p,\b}(\lambda; x_1,\t_1;\cdots,x_{2p},\t_{2p})\\
&=&\sum_{N=1}^{\infty}\sum_{n+n_1+n_2=N}\frac{\l^n}{n!}\frac{(\delta\mu(\lambda))^{n_1}}{n_1!}\int_{{(\L_{\beta})}^{n+n_1}}d^3y_1\cdots d^3y_{n+n_1}\nn\\
&&\int_{{(\L_{\beta})}^{2n_2}} d^3z_{1}\cdots d^3z_{2n_2}\prod_{i,j=1}^{2n_2}\nu(\zz_i-\zz_j)\delta(z_{i,0}-\delta z_{j,0})\nn\\
&&\sum_{\underline{\t}}\Bigg\{\begin{matrix}&\xi^x_1,&\cdots, \xi^x_p,&\xi^z_1,&\cdots, \xi^z_{n_2},&\xi^y_1,\cdots,\xi^y_n\\
&\xi^x_{p+1},&\cdots, \xi^x_{2p},&\xi^z_{n_2+1},&\cdots,\xi^z_{2n_2},&\xi^y_1,\cdots, \xi^y_n
\end{matrix}\Bigg\},\nn
\eea
where $\L_{\beta}=[-\b,\b)\times {\ZZZ^2}$, $N=n+n_1+n_2$ is the total number of vertices, in which $n$ is the number of interaction vertices, to each of which is associated a coupling constant $\lambda$, and $n_1$ is the number of two-point vertices, each one is associated with a bare chemical potential counter-term $\delta\mu(\l)$ and $n_2$ is the number of non-local counter-terms $\nu(\lambda)$. The two-point vertices are also called {\it the counter-term vertices}. We have used Cayley's notation (c.f. \cite{rivbook}) for determinants:
\bea
\Bigg\{\begin{matrix}\xi^x_i\\
\xi^y_j \end{matrix}\Bigg\}=
\Bigg\{\begin{matrix}
x_{i,\t_i}\\ y_{j,\t_j}
\end{matrix}\Bigg\}=\det\Big[\ \delta_{\t_i\t_j}C_{j,\t}(x_i-y_{j})\ \Big].
\eea
It is most conveniently to label the perturbation terms by graphs \cite{RW1}. Before proceeding, let us recall some notations in graph theory.
\begin{definition}\label{defgraph}
Let $I_n=\{1,\cdots,n\}$, $n\in\NNN$, $\cP_n=\{\ell=(i,j), i, j\in I_n, i\neq j\}$ be the set of unordered pairs in $I_n$. A graph $G=\{V_G, E_G\}$ of order $n$ is a set of 
vertices $V_G=I_n$ and edges $E_G\subset\cP_n$, whose cardinalities are denoted by $|V_G|$ and $|E_G|$, respectively. A half-edge, also called an external field, denoted by $(i,\cdot)$, is an object such that each pair of them form an edge: $[(i,\cdot), (j, \cdot)]=(i,j)$, for $i,j\in I_n$. A forest $\cF$ is a graph which contains no loops, i.e. no subset $L=\{(i_1,i_2), (i_2, i_3), \cdots, (i_k, i_1)\}\subset E_{\cF}$ with $k\ge3$. An edge of a forest is called a tree line and an edge in $L$ is called a loop line. A maximally connected component of $\cF$ is called a tree, noted by $\cT$. A tree with a single vertex is allowed. $\cT$ is called a spanning tree if it is the only connected component of $\cF$.
\end{definition}
The perturbation series would be divergent if we fully expand the determinant \cite{RW1}. Instead, we can only partially expand the determinant such that the expanded terms are labeled by forest graphs, which don't proliferate very fast \cite{RW1}.
The set of forests with labeling and satisfy the inclusion relations is called a jungle graph. The canonical way of generating the jungles graphs in perturbation theory is using the BKAR jungle formula (see \cite{AR}, Theorem IV.3):
\begin{theorem}[The BKAR jungle Formula.]\label{ar1}
Let $n\ge1$ be an integer, $I_n=\{1,\cdots, n\}$ be an index set and $\cP_n=\{\ell=(i,j), i, j\in I_n, i\neq j\}$. Let $\cF$ be a forest of order $n$ and $\cS$ be the set of smooth functions from $\RRR^{\cP_n}$ to an arbitrary Banach space. Let ${\bf x}=(x_\ell)_{\ell\in\cP_n}$ be an arbitrary element of $\RRR^{\cP_n}$ and ${\bf 1}\in \RRR^{\cP_n}$ be the vector with every entry equals $1$. Then for any $f\in \cS$, we have:
\be\label{BKAR}
f({\bf 1})=\sum_{\cJ=(\cF_1\subset\cF_1\cdots\subset\cF_{m})\\ m-jungle}\Big(\int_0^1\prod_{\ell\in\cF_m} dw_\ell\Big)\Bigg(\prod_{k=1}^m\Big(\prod_{\ell\in\cF_k\setminus\cF_{k-1}}\frac{\partial}{\partial {x_\ell}}\ \Big)\Bigg)\ f[X^\cF(w_\ell)],
\ee
where the sum over $\cF$ runs over all forests with $n$ vertices, including the empty one which has no edges; $\cJ=({\cF}_0\subset{\cF}_1\cdots\subset{\cF}_m={\cF})$ is a layered object of forests $\{\cF_0,\cdots,\cF_m\}$ labeled by non-negative integers, also called a jungle, in which the last forest $\cF=\cF_{m}$ is a spanning forest of the fully expanded graph $G$ containing $n$ vertices and $2p$ external edges. $\cF_{0}:={\bf V_n}$ is the completely disconnected forest of $n$ connected components, each of which corresponds to the interaction vertex $\VV(\psi,\lambda)$ ( cf. Formula \eqref{potx} ). $X^\cF(w_\ell)$ is a vector ${(x_\ell)}_{\ell\in \cP_n}$ with elements $x_\ell= x_{ij}^\cF(w_\ell)$, which are defined as follows:
\begin{itemize}
\item $x_{ij}^\cF=1$ if $i=j$, 
\item $x_{ij}^\cF=0$ if $i$ and $j$ are not connected by $\cF_k$,
\item $x_{ij}^\cF=\inf_{\ell\in P^{\cF}_{ij}}w_\ell$, if $i$ and $j$ are connected by the forest $\cF_k$ but not $\cF_{k-1}$, where $P^{\cF_k}_{ij}$ is the unique path in the forest that connects $i$ and $j$,
\item $x_{ij}^\cF=1$ if $i$ and $j$ are connected by $\cF_{k-1}$.
\end{itemize}
\end{theorem}
We obtain: 
\bea\label{rexp1}
S_{2p,\b}(\lambda)&=&\sum_{N=n+n_1+n_2} S_{2p,N,\b},\\
S_{2p,N,\b}&=&\frac{1}{n!n_1!n_2!}\sum_{\{{\tau}\}}\sum_{\cJ_{r_{max}},\{\sigma\}}\e(\cJ_{r_{max}})\prod_v\int_{\Lambda_{\beta}} d^3x_v\lambda^n (\delta\mu(\lambda))^{n_1} (\nu)^{n_2}\nn\\
&&\quad\quad\cdot\prod_{\ell\in\cF}\int dw_\ell
C_{r,\sigma_\ell}(x_\ell,x'_\ell){\det}[ C_{r,\sigma,\tau}(w)\ ]_{left}\ .\label{rexp11}
\eea
where the sum is over the jungles $\cJ'_{r_{max}}=({\cF}_0\subset{\cF}_1\cdots\subset{\cF}_{r_{max}})$, in which each layered forest is labeled by the scale index $r$, $r=0,1,\cdots, r_{max}$; $\e(\cJ_{r_{max}})$ is a product of the factors $\pm1$ along the jungle. The term
$\det[C_{r,\sigma}(w)]_{left}$ is the determinant for the remaining $2(n+1)\times 2(n+1)$ dimensional square matrix, which has the same form as \eqref{rexp1}, but is multiplied by the interpolation parameters $\{w\}$. So it is still a Gram matrix. This matrix contains all the unexpanded fields and anti-fields that don't form tree propagators. Let $j_f$ be the index of a field or anti-field $f$, then the $(f,g)$ entry of the determinant reads:
\bea\label{intc1}
C_{r,\sigma,\tau}(w)_{f,g}=\delta_{\t(f)\t'(g)}\sum_{v=1}^n\sum_{v'=1}^n\chi(f,v)\chi(g,v') x^{\cF,r_f}_{v,v'}(\{w\})C_{r,\s(f),\t(f)}(x_v,x_{v'}),
\eea
where $[x^{\cF,r_f}_{v,v'}(\{w\})]$ is an $n\times n$ dimensional positive matrix, whose elements are defined in the same way as in \eqref{BKAR}:
\begin{itemize}
\item If the vertices $v$ and $v'$ are not connected by $\cF_r$, then $x^{\cF,r_f}_{v,v'}(\{w\})=0$,
\item If the vertices $v$ and $v'$ are connected by $\cF_{r-1}$, then $x^{\cF,r_f}_{v,v'}(\{w\})=1$,
\item If the vertices $v$ and $v'$ are connected by $\cF_r$ but not $\cF_{r-1}$, then  $x^{\cF,r_f}_{v,v'}(\{w\})$ is equal to the infimum of the $w_\ell$ parameters for $\ell\in\cF_r/\cF_{r-1}$ which is in the unique path connecting the two vertices. The natural convention is that $\cF_{-1}=\emptyset$ and that $x^{\cF,r_f}_{v,v'}(\{w\})=1$.
\end{itemize}
\begin{remark}\label{ctbd}
Since the counter-terms only contribute the factors $|\delta\mu|^{n_1}$ and $|\nu|^{n_2}$ to the amplitudes of the correlation functions and since these factors are bounded (cf.  Theorem \ref{flowmu} and \ref{mainc}), we will forget them and replace them by constants, just for simplicity. We will retain them when we study the bounds for $|\delta\mu|$ and $|\nu|$.
\end{remark}

Taking the logarithm on $S_{2p,\b}$, we obtain the {\it connected} $2p$-point Schwinger function $S^c_{2p,\b}$:
\bea\label{rexp2}
S^c_{2p,\b}&=&\sum_{N=n+n_1+n_2}S^c_{2p,N,\b},\\
S^c_{2p,N,\b}&=&\frac{c^N\lambda^n}{n!n_1!n_2!}\sum_{\{{\tau}\}}\sum_{\cJ'_{r_{max}}}\epsilon (\cJ'_{r_{max}})\prod_v\int_{\L_{\b}} d^3x_v\prod_{\ell\in\cT}\int dw_\ell C_{r,\sigma_\ell}(x_\ell,x'_\ell){\det}[ C_{r,\sigma,\tau}(w)]_{left}\nn ,\\\label{rexp3}
\eea
in which $S^c_{2p,\b}$ has almost the same structure as $S_{2p,\b}$, except that the summation over jungles is restricted to the ones $\cJ'_{r_{max}}=({\cF}'_0\subset{\cF}'_1\cdots\subset{\cF}'_{r_{max}}={\cT})$, in which the final layered forest is a spanning tree $\cT$ of $G$. Without losing generality, suppose that a forest $\cF_r$ contains $c(r)\ge1$ trees, denoted by $\cT_r^k$, $k=1,\cdots, c(r)$. A link of $\cT_r^k$ is noted by $\ell(T)$. To each $\cT_r^k$ we introduce an extended graph (cf. Definition \ref{defgraph}) $G^k_r$, which contains $\cT_r^k$ as a spanning tree and contains a set of half-edges, $e(G^k_r)$, such that the cardinality of $e(G^k_r)$, denoted by $|e(G^k_r)|$, is an even number. By construction, the scale index of any external field, $r_f$, is greater than $r_{\ell(T)}$, and each connected component $G^k_r$ is contained in a unique connected component with a lower scale index $r$. The inclusion relation of the graphs $G^k_r$, $r=0,\cdots,r_{max}$, has a tree structure, called the Gallavotti-Nicol\`o tree.

\begin{figure}[htp]
\centering
\includegraphics[width=0.45\textwidth]{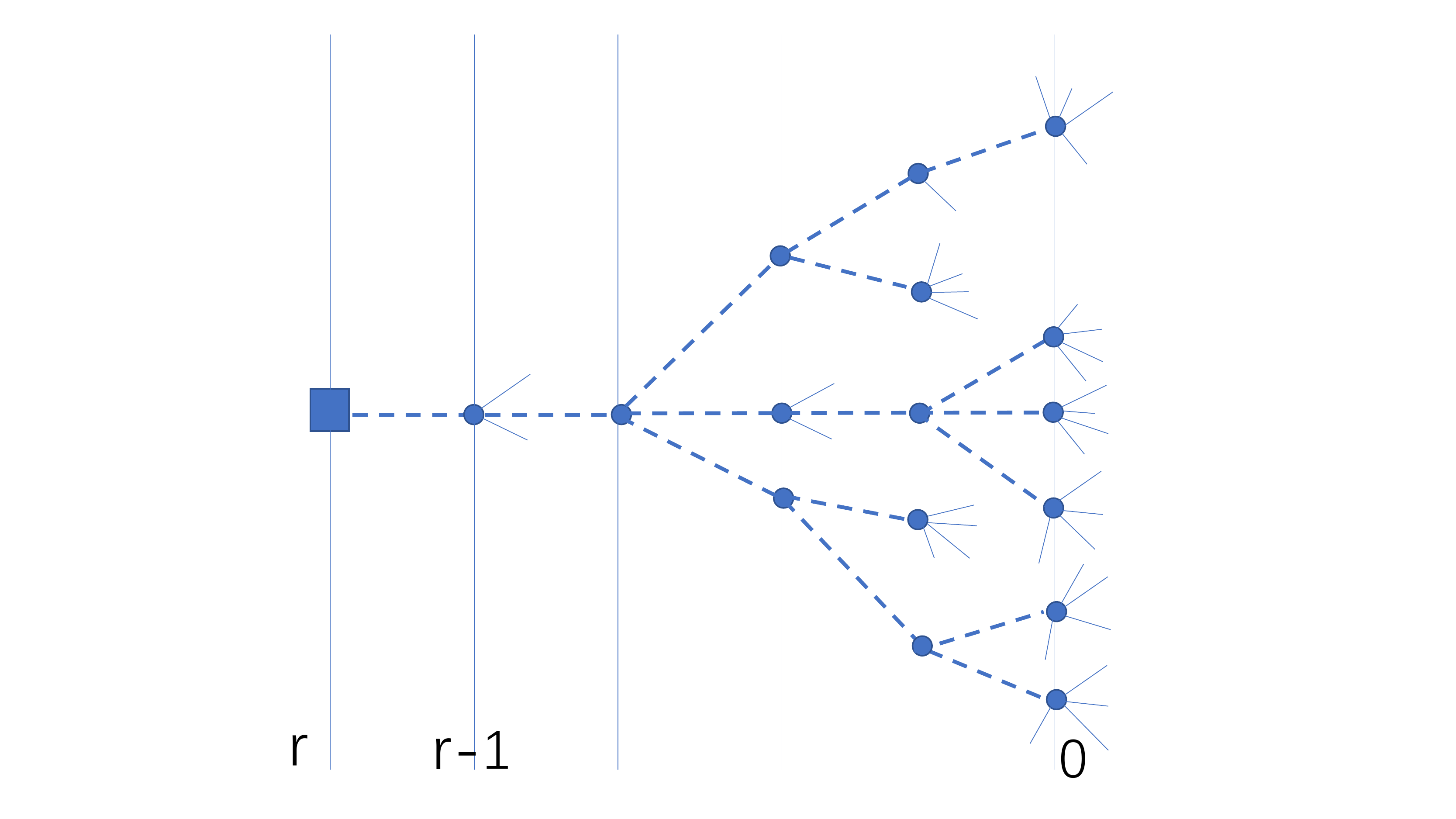} \includegraphics[width=0.46\textwidth]{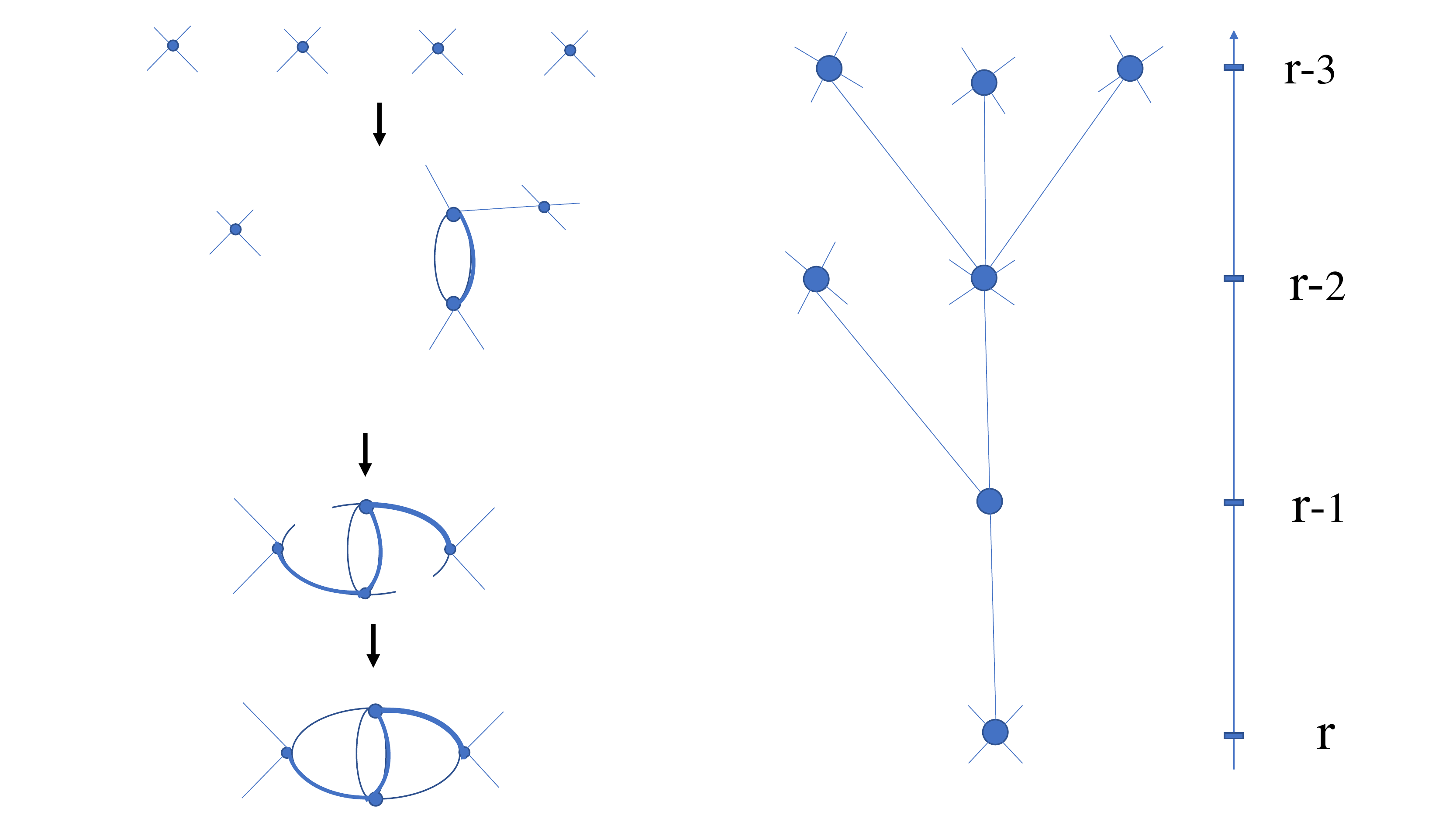}
\caption{\label{gn1}A Gallavotti-Nicol\`o tree (l.h.s.) and the grouping of the quadruped Feynman graphs into a GN tree (r.r.s.). The round dots
are the nodes and bare vertices, the square is the root and thin lines are the external fields. The dash lines indicate the inclusion relations.
}
\end{figure}

\begin{definition} [cf. \cite{GN}]
A Gallavotti-Nicol\`o tree (GN tree for short) $\cG^{r_{max}}$ (see Figure \ref{gn1} for an illustration) is an abstract tree graph in which the vertices, also called the nodes, correspond to the extended graphs $G^k_r$, $r\in[0,r_{max}]$, $k=1,2,\cdots, c(r)$ and the edges are the inclusion relations of these nodes. The node $G_{r_{max}}$, which corresponds to the full Feynman graph $G$, is called the root of $\cG^{r_{max}}$. Obviously each GN tree has a unique root. Elements of the set ${\bf V_N}=\cF_0$ are called the bare nodes or leaves and $|\cF_0|$ is called the order of $\cG^{r_{max}}$. A GN tree of order $N$ is denoted as $\cG^{r_{max}}_N$.
\end{definition}

We can rewrite \eqref{rexp2} as
\bea\label{s2pa}
S^c_{2p,\b} &=&\sum_{n}S^c_{2p,n,\b}\lambda^n,\\
S^c_{2p,n,\b}&=&{ \frac{1}{n!}}{\frac{1}{n_1!}}{\frac{1}{n_2!}}\sum_{ \{{\t}\},\cG^{r_{max}}, \cT} \ 
\sum_{\cJ'_{r_{max}},\{\sigma\}}
\epsilon (\cJ'_{r_{max}}) \prod_{j=1}^{n}   \int d^3x_{j}  \delta(x_1)
\prod_{\ell\in \cT} \int_{0}^{1} dw_{\ell}
C_{\t_{\ell},\si_{\ell}} 
(x_{\ell}, \bar x_{\ell}) \nonumber\\
&&\quad \prod_{i=1}^{n} 
\chi_{i}(\si)    
[\det C_{r,\sigma,\tau}(w)]_{{ left}} \ .\label{form} 
\eea
By \eqref{intc1}, each matrix element of $\det[C_{j,\sigma}(w)]_{left}$ can be written as an inner product of two vectors:
\be
C_{r,\sigma}(w)_{f,g;\tau,\tau'}=( e_\tau\otimes A_f(x_v, ),  e_{\tau'}\otimes B_g(x_{v'},))\ ,
\ee
in which the unit vectors $e_{\uparrow}=(1,0)$, $e_{\downarrow}=(0,1)$ are the spin variables,
$A_f$, $B_g$ are complex functions in a certain Hilbert space, and the inner product is defined as:
\be
( e_\tau\otimes A_f(x, \cdot),  e_{\tau'}\otimes B_g(y,\cdot))= 
(e_\tau\cdot e_{\tau'})\int dz\bar A_f(x, z)B_g(y, z).
\ee
So we have 
\bea
A_f&=&\sum_{k\in\DD_{\beta, \L}}\sum_{v=1}^n\chi(f,v)[x^{\cF,r_f}_{v,v'}(\{w\})]^{1/2}e^{-ik\cdot x_v}\cdot\\
&&\quad\quad\quad\cdot\Big[\ \chi_r[k_0^2+e_0^2(\kk,\mu_0)]\cdot
v_{s_+}[q^2_+]\cdot v_{s_-}[q^2_-]\ \Big]^{1/2},\nn
\eea
\bea
B_g&=&\sum_{k\in\DD_{\beta, \L}}\sum_{v'=1}^n\chi(g,v')[x^{\cF,r_f}_{v,v'}(\{w\})]^{1/2}e^{-ik\cdot x_{v'}}\hat C(k)_{\tau_\ell,\sigma_\ell}\\
&&\quad\quad\quad\cdot\Big[\ \chi_r[k_0^2+e_0^2(\kk,\mu_0)]\cdot
v_{s_+}[q^2_+]\cdot v_{s_-}[q^2_-]\ \Big]^{1/2}\nn,
\eea
which satisfy
\bea
\Vert A_f\Vert^2_{L^2}\le K\g^{-j_f/2-s_{f,+}/2-s_{f,-}/2},\quad \Vert B_f\Vert^2_{L^2}\le K\g^{j_f/2-s_{f,+}/2-s_{f,-}/2},
\eea
for some positive constant $K$. By Gram-Hadamard's inequality \cite{Le, GK} and using the relation $l_f=s_{f,+}+s_{f,-}-3j/2+j_0/2$, 
we have
\be
\det\nolimits_{{ left}}\le K^n\prod_{f\ left}\g^{-(l_f/2+3j_f/4-j_0/4)}
\ee 
for some positive constant $K$ that is independent of $r$ and $l$. In order to encode the constraints from the conservation of momentum, we introduce an indicator function $\chi_{i}(\{\si\})= \chi (\sigma^1_i, \sigma^2_i, \sigma^3_i, \sigma^4_i)$ at each vertex in the graph, defined as follows: $\chi_{i}(\{\si\})$ equals to $1$ if the sector indices $\{\si\}$ satisfy the constraints in Lemma \ref{secmain}, and equals to $0$ otherwise. Integrating over the all position variables except the fixed one,
$x_1$, we obtain the following bound:
\bea 
|S^c_{2p,n,\b} | &\le& {\frac{K^n}{n!}}\sum_{ \cG^{r_{max}}, \cT} \sum'_{\{\si \}} 
\prod_{i=1}^{n} \chi_{j}(\si) 
\prod_{\ell \in \cT} \g^{2r_{\ell}}
\prod_{f} \g^{-r_f/2-l_f/4}\g^{-(j_f-j_0)/8}\nn\\
&\le&  {\frac{K^n}{n!}}\sum_{\cG^{r_{max}}, \cT} \sum'_{\{\si \}} 
\prod_{i=1}^{n} \chi_{j}(\si) 
\prod_{\ell \in \cT} \g^{2r_{\ell}}
\prod_{f} \g^{-r_f/2-l_f/4} \label{absol1}
\eea
in which the last product runs over all the $4n$ fields
and anti-fields, and the summation $\sum'$ means that we have taken into account the constraints on the sector indices among the different connected components in the GN tree. We have also used the fact that $j_0\le j_f$. We have the following lemma concerning the last two terms in \eqref{absol1}.
\begin{lemma}\label{indmain}
Let $c(r)$ be the number of connected components at level $r$ in the GN tree, we have the following inductive formulas:
\bea &&\prod_{f} \g^{-r_{f}/2}= \prod_{r=0}^{r_{max}}\ \prod_{k=1}^{c(r)} \g^{-|e(G_r^k)|/2}\ ,
\label{induc1}\\
&&\prod_{\ell \in \cT} \g^{2r_{\ell}}=\g^{-2r_{max}-2}
\prod_{r=0}^{r_{max}}\ \prod_{k=1}^{c(r)}  \g^{2}\ .
\label{induc2}
\eea
\end{lemma}
\begin{proof}
See the proof of Lemma $4.1$ in \cite{RWang1}.
\end{proof}
\begin{theorem}[The power counting theorem]\label{tpc}
There exists a positive constant $K$ which may depend on the model but is independent of the scale indices, such that the connected, $2p$-point Schwinger functions, with $p\ge0$, satisfy the following bound: 
\be\label{pc1} | S^c_{2p,n,\b} | \le {\frac{K^n}{n!}}
\sum_{ \cG^{r_{max}}, \cT}\ \sum'_{\{\si \}}\prod_{i=1}^{n}\ \big[\chi_{i}(\si)\g^{-(l_i^1 + l_i^2 + l_i^3 + l_i^4)/4}\big]\ 
\prod_{r=0}^{r_{max}}\prod_{k}  \g^{2-e(G_r^k)/2}\ .
\ee
So the two-point functions are relevant, the four point functions are marginal and the Schwinger functions with external legs $2p\ge6$ are irrelevant.
\end{theorem}

\begin{proof}
Writing the product $\prod_f\g^{-l_f/4}$ in Formula \eqref{absol1} as $\prod_{i=1}^{n}e^{-(l_i^1 + l_i^2 + l_i^3 + l_i^4)/4}$, using Lemma \ref{indmain} and conservation of momentum at each vertex $i$, the result follows.
\end{proof}

\subsection{The sector counting lemma}
Now we consider summation over the sector indices in \eqref{pc1}, which is easily getting unbounded if we don't take into account the constraints placed by the conservation of momentum. We have the following lemma for the sector counting problem.
\begin{lemma}[Sector counting lemma for a single bare vertex]\label{sec1}
Let the four half-fields attached to a vertex be $f_1,\cdots, f_4$ with scale indices $j_1,\cdots, j_4$, sector indices $\si_1=(s_{1,+}, s_{1,-}),\cdots,\si_4=(s_{4,+}, s_{4,-})$. Let the generalized scale indices $r_1\cdots, r_4$ associated to the four fields be assigned as
$r_{f_1}=r_{f_2}=r_{f_3}=r$ and $r_{f_4}>r$. Then there exists a positive constant $K$, which is independent of the scale indices and the sector indices, such that for fixed $\si_4$, we have
\be\label{ss1}
\sum_{\si_1, \si_2, \si_3} \chi (\si_1, \si_2, \si_3, \si_4) 
\gamma^{-(l_1+l_2+l_3 )/4} \le K(j+j_0)\ .\ee
\end{lemma}
\begin{proof}
Let $f_1,\cdots, f_4$ be the four fields hooked to a vertex $i$ in a node $G_r^k$ with sector indices $\s_1,\cdots,\s_4$ and depth indices $l_i^1,\cdots, l_i^4$, respectively. Among the four fields, we can always choose one, say, $f_4$, as the {\it root} field, for which the index $r_4$ is greater than all the other indices $r_1=r_2=r_3=r$. We can always organize the sector indices $\s_1=(s_{1,+}, s_{1,-}),\cdots, \s_3=(s_{3,+}, s_{3,-})$ such that $s_{1,+}\le s_{2,+}\le s_{3,+}$ and $s_{1,-}\le s_{2,-}\le s_{3,-}$. Then, by Lemma \ref{secmain}, either $\s_1$ collapses with $\s_2$, or one has $s_{1,+}=s_{1,-}=j_1$, such that $j_1<\min\{j_2,\cdots,j_4\}$. 
So we only need to consider the following cases:
\begin{itemize}
\item if $\sigma_1\simeq\sigma_2$, then $s_{2,+}=s_{1,+}\pm1$ and $s_{2,-}=s_{1,-}\pm1$. The depth indices are arranged as $l_1\le l_2\le l_3$. Then the l.h.s. of \eqref{ss1} is bounded by
\be
\sum_{\si_1, \si_3}
\g^{-(2l_1+l_3 )/4}= \sum_{\si_1}\g^{-l_1/2}\sum_{\si_3}
\g^{-l_3 /4}\ .\nn
\ee
Using the fact that $2r_k=j_k+s_{k,+}+s_{k,-}$ and $l_k=s_{k,+}+s_{k,-}-3j_k/2+j_0/2$, we obtain
\be l_k=\frac52(s_{k,+}+s_{k,-})-3r_k+j_0/2,  k=1,\cdots,3.\label{indexre}\ee 
For fixed $s_1=(s_{1,+},s_{1,-})$, summation over $\s_3=(s_{3,+},s_{3,-})$ can be easily bounded by:
\bea
&&\sum_{\si_3=(s_{3,+},s_{3,-})}\g^{-l_3 /4} = \sum_{\si_3=(s_{3,+},s_{3,-})} \g^{-(l_3-l_1) /4}
\g^{-l_1/4}\nn\\
&&\le \sum_{s_{3,+}\ge s_{1,+}}\g^{-5(s_{3,+}- s_{1,+})/8}\sum_{s_{3,-}\ge s_{1,-}}\g^{-5(s_{3,-}-s_{1,-})/8}\g^{-l_1/4} \le K_1\cdot  \g^{-l_1/4}\ ,\nn
\eea
for some positive constant $K_1$ which is independent of the scale indices.
Now we consider the summation over $\s_1$. By the constraint $s_{1,+}+ s_{1,-}\ge 6r/5-j_0/5$ and take into account the factor $\g^{-l_1/4}$ from the above formula, we have:
\bea
\sum_{\si_1}\g^{-l_1/2}\cdot \g^{-l_1/4}&=&\sum_{s_{1,+}}\g^{-3s_{1,+}/4}\sum_{s_{1,-}=3j/2-j_0/2-s_{1,+}}^{j}\g^{-3s_{1,-}/4+9j/8-3j_0/8}\nn\\
&\le&\sum_{s_{1,+}=\max(0,(j-j_0)/2)}^{j}1.\label{secsum}
\eea
If $j\le j_0$, then \eqref{secsum} is bounded by $j_0$. If $j\ge j_0+1$, \eqref{secsum} is bounded by $(j+j_0)/2$. Since we are mainly interested in the case of $j\gg j_0$, we have
\be
\sum_{\si_1, \si_3}
\g^{-(2l_1+l_3 )/4}\le (j+j_0)/2.
\ee
Choosing $K=K_1/2$, we proved the lemma for this case.
\item if $j_1=s_{1,+}=s_{1,-}$ and is the smallest index among the four scale indices, summing over $\s_1$ is simply bounded by $1$ (since $s_{1,\pm}$ are fixed and there is no summation) and summing over $\s_3\ge \s_2$ is bounded by a constant. Finally the sum over $\s_2$ is bounded by $K(j+j_0)$, for some positive constant $K$ that is independent of the scale indices.
\end{itemize}
\end{proof}
\section{The Convergent Schwinger Functions}
\subsection{More notations about the Gallavotti-Nicol\`o trees}
Before proceeding, let us introduce the following notations concerning some specific Gallavotti-Nicol\`o trees. Let $G_r^k$ be a node in a GN tree $\cG^{r_{max}}$ with root scale index $r_{max}$ and $e(G_r^k)$ be the set of external fields of $G_r^k$, whose cardinality is denoted by $|e(G_r^k)|$. A biped $b$ is a node in $\cG^{r_{max}}$ such that $|e(G_r^k)|=2$ and a quadruped $Q$ is a node in $\cG^{r_{max}}$ such that $|e(G_r^k)|=4$. The set of all bipeds is denoted by $\cB:=\{G_r^k,\ r=1,\cdots, r_{max}; k=1,\cdots, c(r)\ \big|\ |e(G_r^k)|=2\}$. Similarly, the set of all quadrupeds in a $\cG^{r_{max}}$ is noted by $\cQ$.
\begin{definition}
A biped tree $\cG^{r_{max}}_\cB$ is defined as a subgraph of a Gallavotti-Nicol\`o tree in which the set nodes, noted by $V(\cG^{r_{max}}_\cB)$, consists of the following elements: i) the bare nodes $\VV$ of $\cG^{r_{max}}$, ii), the bipeds $b$ and iii) the root node which corresponds to the complete graph $G$. The edges of $\cG^{r_{max}}_\cB$ are the natural inclusion relations for the nodes in $V(\cG^{r_{max}}_\cB)$. The set of external fields of a biped $b$ is noted by $e_b=\{\bar\psi_b,\psi_b\}$, and the set of external fields of $\cG^{r_{max}}_\cB$ is noted by ${\cal EB}$. We have ${\cal EB}:=\big(\cup_{b\in B}\ e_b\big)\setminus e(G)$, where $e(G)$ is the set of external fields of the complete graph $G$.  A quadruped tree $\cG^{r_{max}}_\cQ$ is defined as a subgraph of $\cG^{r_{max}}$ whose set of nodes, noted by $V(\cG^{r_{max}}_\cQ)$, composes of the following elements: the bare nodes of $\cG^{r_{max}}$, the quadruped $\cQ$ and the root of $\cG^{r_{max}}$ which is the complete graph $G$. The edges of $\cG^{r_{max}}_\cQ$ are the inclusion relations of its nodes. The set of external fields associated to $q$ is noted by $e_q$, and the set of external fields of $\cQ$ is noted by ${\cal EQ}$. We have ${\cal EQ}=(\cup_{Q\in\cQ}e_Q)\setminus e(G)$. 
\end{definition}

\begin{definition}\label{gn3}
A convergent Gallavotti-Nicol\`o tree $\cG^{r_{max}}_{\cC}$ is a subgraph of a Gallavotti-Nicol\`o tree whose nodes form the set $V(\cC):=\big\{ G_{r}^{k},\ r=1,\cdots, r_{max},\ k=1\cdots c(r)\big\vert\ |e( G_{r}^{k})|\ge 6\big\}$ and the edges are the natural inclusion relations of the nodes.
%
\end{definition}
We have the following definitions for the corresponding Schwinger functions.
\begin{definition}
The connected Schwinger functions $S^c_{\cC,2p,\b}$, $p\ge3$, related to the convergent GN trees $\{\cG^{r_{max}}_{\cC}\}$ are called the convergent Schwinger functions. Similarly, the connected Schwinger functions $S^c_{\cQ,\b}$ related to the quadrupeds $\{\cG^{r_{max}}_\cQ\}$ are called the quadruped Schwinger functions, and the connected Schwinger functions $S^c_{\cB,\b}$, which correspond to the biped GN trees $\{\cG^{r_{max}}_{\cB}\}$, are called the biped Schwinger functions.
\end{definition}
\subsection{The $2p$-point Schwinger functions with $p\ge3$.}
The perturbation series of $S^c_{\cC,2p}$ can be written as
\be
S^c_{\cC,2p,\b}(\lambda)=\sum_n\l^n S_{\cC,2p,n},\ee
\be\label{conv1}
S_{\cC,2p,n} = {\frac{1}{n!}}\sum_{{\cal B} = \emptyset,
{\cal Q}=\emptyset,  \{\tau\}, \cJ'_{r_{max}} }
\sum'_{\{\si \}} \ep (\cJ'_{r_{max}})\prod_{v} \int_{\Lambda_{\beta}} dx_{v} 
\prod_{\ell\in \cT} \int_{0}^{1} dw_{\ell}
C_{r,\si_{\ell}} (x_{\ell}, y_{\ell})
[\det C(w)]_{{ left}}.
\ee
We have the following theorem:
\begin{theorem}[The Convergent contributions (see also \cite{RWang1}]\label{cth1}
There exists a positive constant $c_1$ independent of the scale index such that the connected Schwinger functions $S^c_{\cC, 2p,\b}(\lambda)$, $p\ge3$, are analytic functions of $\lambda$, for $\l\in\RR^{\cC}_T:=\{\l\in\RRR,\ |\lambda\log\mu_0 T|\le c_1\} $.
\end{theorem}
\begin{proof}
The proof follows closely \cite{RWang1}. Formula \eqref{conv1} can be written as:
\bea\label{conv2}
S_{\cC,2p,n} &=& {\frac{1}{n!}}\sum_{\{G^k_r, {r=0},\cdots,r_{max}; {k=1},\cdots,c(r)\}, { {\cal B} = \emptyset,
{\cal Q}=\emptyset}}\sum_{\tau}
\sum_{\{\si \}}' \ep (\cJ'_{r_{max}})\nn\\
&&\quad\quad \prod_{v} \int_{\Lambda_{\beta}} dx_{v} 
\prod_{\ell\in \cT} \int_{0}^{1} dw_{\ell}
C_{r,\si_{\ell}} (x_{\ell}, y_{\ell})
[\det C(w)]_{left},
\eea
in which we make the sum over $\cJ'_{r_{max}}$ more explicit.
By Theorem \ref{tpc}, we have:
\bea\label{conv3}
|S_{\cC,2p,n}| &\le& 
{\frac{K^n}{n!}}
\sum_{\{G^k_r, {r=0},\cdots,r_{max}; {k=1},\cdots,c(r)\},{ {\cal B} = \emptyset,
{\cal Q}=\emptyset}}\sum_{\underline\tau,\cT} \sum'_{\{\si \}}\prod_{i=1}^{n}\ \Big[\ \chi_{i}(\si)e^{-(l_{i,1} + l_{i,2} + l_{i,3} + l_{i,4})/4}\ \Big]\nn\\ 
&&\quad\cdot\prod_{i=1}^n \g^{-[r_{i,1} + r_{i,2} + r_{i,3} + r_{i,4}]/6}
\label{cpt1},
\eea
in which we have used the fact that $2-|e(G^k_r)|/2\le-|e(G^k_r)|/6$, for $|e(G^k_r)|\ge 6$, and the fact that $$\prod_{r=0}^{r_{max}}\g^{-|e(G^k_r)|/6}=\prod_{i=1}^n \g^{-[r_{i,1} + r_{i,2} + r_{i,3} + r_{i,4}]/6}.$$
Now we consider the summation over the sector indices. Without losing generality, we can choose the field $f_4$ as the root field.
The constraint from conservation of momentum implies that either the two smallest sector indices among the four, chosen as $s_{1,+}$ and $s_{2,+}$, are equal (modulo $\pm1$), or the smallest sector index $s_{1,+}$ is equal to the $j_1$, which is the smallest scale index.
Then summing over the sector indices $(s_{1,+},s_{1,-})$ results in a factor $\max \{r_{i,1}, r_{i,2}\}$. Now we consider the summation over the sectors $(s_{3,+},s_{3,-})$, for which we obtain (see the proof of Lemma \ref{sec1})
\be
\sum_{s_{3,+},s_{3,-}}\g^{-l_{i,3}/4}\le K_1.(j_{i,3}+j_0)/2,
\ee
for some positive constant $K_1$. The total summation is bounded by $K_1 \bar r\cdot r_{i,3}$, where $\bar r=\max\{r_{i,1}, r_{i,2}\}$. Summation over the sectors which are not the root sectors is bounded by
\be
\prod_{i=1}^n\sum_{r_i^1,\cdots, r_i^{4}=0}^{r_{max}}[\bar r \g^{-r_{i,1}/6}\g^{-r_{i,2}/6}\cdot r_{i,3}\g^{-r_{i,3}/6}\g^{-r_{i,4}/6}]\le K_2.
\ee 
Finally, summation over the sector indices for the root fields, one for each vertex $i$, is bounded by 
\be K(j_{max}+j_0)=K(|\log T|+\vert\log \mu_0\vert),\label{fscale}\ee 
for some positive constant $K$ that is independent of $j$ and $j_0$. Summing over all the GN trees and
spanning trees cost a factor $K_3^n n!$, where $K_3$ is certain positive constant (see \cite{Riv04} for the detailed proof of this combinatorial result.). Choosing the positive constant $C=K\cdot K_1\cdot K_2\cdot K_3$, we have
\be
|S_{\cC,2p,n}|\le \sum_{n=0}^\infty C^n|\lambda|^n|\log \mu_0T|^n.
\ee
Obviously the above series is convergent for $C|\lambda| |\log \mu_0T|<1$. Choosing $c_1\le1/C$
and define
\be
\RR^{\cC}_T=\{\l\in\RRR,\ |\lambda\log\mu_0 T|\le c_1\},
\ee
we conclude this theorem.
\end{proof}

\subsection{The quadruped Schwinger functions}
Now we consider the bound for the quadruped Schwinger functions $S^c_{\cQ,\b}$. To take into account the conservation of moment, we introduce an indicator function $\chi_i(\{\sigma\})$ to each bare vertex. We introduce also an indicator function $\chi_Q(\{\sigma\})$
to each quadruped $Q$, which is defined as follows: $\chi_Q(\{\sigma\})$ is equal to $1$ if the sector indices $\{\sigma\}=\{\s_{Q,1},\cdots,\s_{Q,4}\}$ of the external fields of $Q$ satisfy the constraints in Lemma \ref{secmain}, and is equal to $0$ otherwise. Then the quadruped Schwinger functions can be written as
\bea\label{conv2q}
S^c_{\cQ,\b}(\lambda)&=&\sum_{n=0}^\infty \lambda^n S_{\cQ,n},\\
S_{\cQ,n} &=& {\frac{1} {n!}}\sum_{\cG^{r_{max}}_\cQ,\cal{EQ}}\sum_{\underline\tau}
\sum_{\{\si \}}' \ep (\cJ'_{r_{max}})\prod_{i=1}^n\chi_i(\{\sigma\})\prod_{Q\in\cQ}\chi_Q(\{\sigma\})\nn\\
&&\quad\quad \prod_{v} \int_{\Lambda_{\beta}} dx_{v} 
\prod_{\ell\in \cT} \int_{0}^{1} dw_{\ell}
C_{r_\ell,\si_{\ell}} (x_{\ell}, y_{\ell})
[\det C(w)]_{left}.
\eea

\begin{definition}[The maximal sub-quadruped]
Let $Q\in\cG^{r_{max}}_\cQ$ be a quadruped which is not a leaf. By the Gallavotti-Nicol\`o tree structure of $\cG^{r_{max}}_\cQ$, $Q$ must be linked directly to a set of quadrupeds $\{Q'_1,\cdots,Q'_{d_Q}\}$, $d_Q\ge1$, called the maximal sub-quadrupeds of $Q$. These sub-quadrupeds could be either the bare vertices or some general quadrupeds.
\end{definition}
\begin{remark}
A maximal sub-quadruped $Q'$ of $Q$ may still contain some sub-quadrupeds $Q''_1,\cdots, Q''_{d(Q')}$. The inclusion relation between $Q$ and $Q''$ is not an edge of the quadruped Gallavotti-Nicol\`o tree.
\end{remark}
We have the following theorem:
\begin{theorem}\label{mqua}
Let $T=1/\b>0$ be the temperature and $S^c_{\cQ,\b}(\lambda)$ be the quadruped Schwinger function. There exists a positive constant $c$, which is independent of the scale indices, such that the perturbation series for the quadruped Schwinger function is convergent in the domain $\{\lambda\in\RRR,\ |\lambda|<c/|\log \mu_0T|^2\}$.
\end{theorem}
In order to prove this theorem, we shall we consider first the summation over sector indices for a quadruped, for which we have the following lemma. 
\begin{lemma}[Sector counting lemma for quadrupeds]\label{secqua}
Let $Q$ be a quadruped of scale indices $r$, which is linked to $d_Q$ maximal sub-quadrupeds ${Q'_1,\cdots,Q'_{d_Q}}$. Let the external fields of $Q$ be ${f_{Q,1},\cdots,f_{Q,4}}$, with scale indices ${r_{Q,1},\cdots,r_{Q,4} }$ and sector indices ${\s_{Q,1},\cdots,\s_{Q,4}}$, respectively. Let the external fields of $Q'_v$, $v=1,\cdots Q_d$, be ${f_{v,1},\cdots,f_{v,4} }$, with scale indices ${r_{v,1},\cdots,r_{v,4} }$ and sector indices $\s_{v,1}=(s_{v,1,+},s_{v,1,-})$, $\cdots$, $\s_{v,4}=(s_{v,4,+},s_{v,4,-})$, respectively. Let $\chi_v(\{\sigma_v\})$ be the characteristic function at the sub-quadruped $Q'_v$, with $v=1,\cdots, d_Q$, we have
\be
\sum_{\{\sigma_1\},\cdots,\{\sigma_{d_Q}\}} \prod_{v=1}^{d_Q}\chi_v(\{\sigma_v\})\chi_Q(\{\sigma\})e^{-[l_{v,1}+l_{v,2}+l_{v,3}+l_{v,4}]/4}\le K_1^{d_Q-1} {(j_v+j_0)}^{d_Q-1},\label{qsecsum}
\ee
for some positive constant $K$.
\end{lemma}

\begin{proof}
The proof is almost identical to the proof of Lemma $5.1$ in \cite{RWang1}, except that the sum over the sector indices for the root field in Lemma $5.1$ in \cite{RWang1} should be bounded by $\sum_{(s_{v,+},s_{v,-})}\g^{-l/4 }\le K_1.(j_v+j_0)/2$. We shall skip the proof and ask the reader to look at
\cite{RWang1} for details.
\end{proof}
\begin{proof}[Proof of Theorem \ref{mqua}]
In order to sum over all the quadruped trees, it is useful to keep the GN tree structure explicit and write a quadruped tree as $\cG^{r_{max}}_\cQ=\{Q_r^k, r=0,\cdots, r_{max}, k=1,\cdots, c(r)\}$. The quadruped Schwinger functions satisfy the following bound (cf. Formula \eqref{cpt1}):
\bea
|S_{\cQ,n}| &\le& 
{\frac{K^n}{n!}}
\sum_{\{Q^k_r, {r=0},\cdots,r_{max}; {k=1},\cdots,c(r)\},{ {\cal B} = \emptyset}}\sum_{\underline\tau,\cT} \sum'_{\{\si \}}\prod_{Q\in\cQ}\chi_Q(\{\sigma\})\nn\\&&\cdot\prod_{i=1}^{n}\ \Big[\chi_{i}(\si)e^{-[l_{v,1}+l_{v,2}+l_{v,3}+l_{v,4}]/4}\Big]
\prod_{r=0}^{r_{max}}\g^{2-|e(G^k_r)|/2}\ .\label{convq}
\eea
Summation over the sector indices is to be performed from the leaves of a quadruped tree to the root. The first quadruped $Q_1$ that we encounter contains only the bare vertices as the maximal sub-quadrupeds. The second quadruped $Q_2$ contains $Q_1$, some other quadrupeds at the same scale index than $Q_1$, and bare vertices. More quadruped will be encountered when we are going towards the root. In this process, we will apply Lemma \ref{secqua} to each quadruped $Q$ that we meet, until we arrive at the root node of $\cG^{r_{max}}_\cQ$. Then there exists a constant $K_2$ independent of the scale index such that:
\bea
&&|S_{\cQ,n}| \le \prod_{Q\in\cQ}\Big[K_2^{d_Q} \sum_{r=0}^{r_{max}} {(j+j_0)}^{d_Q-1}\Big]\nn\\
&&=\prod_{Q\in\cQ}\Big[K_2^{d_Q} \sum_{j=0}^{\frac32j_{max}} {(j+j_0)}^{d_Q-1}\Big]
\le \prod_{Q\in\cQ}K_3^{d_Q}|\log \mu_0T|^{d_Q},
\label{convq2}
\eea
in which $K_3$ is another positive constant independent of the scale index. In \eqref{convq2} we have used the fact that $j_0=\EEE(\log{1/\mu_0})$ and $j_{max}=\EEE(\log 1/T)$, and the fact that the number of quadruped trees with $n$ vertices is bounded by $c^n n!$ (see \cite{Riv04}), which is a variation of Cayley's theorem concerning the number of labeled spanning trees with fixed vertices.
Using following well-known inductive formula 
\be\label{ind4p}\sum_{Q\in \cQ} d_Q=|\cQ|+n-1\le 2n-2,
\ee 
where $| \cQ|$ is the cardinality of the set of quadrupeds $\cQ$, for which we have $| \cQ|\le n-1$,
we can prove the following bound:
\be
|S_{\cQ,n}|\le\prod_{Q\in  \cQ}K_3^{d_Q}|\log \mu_0T|^{d_Q}\le K_3^{2n-2}|\log\mu_0T|^{2n-2},
\ee
and
\be
|S^c_{\cQ,\b}(\lambda)|\le\sum_{n=0}^\infty K_3^{2n-2}\cdot\lambda^n\cdot|\log\mu_0T|^{2n-2}.
\ee
Let $0<c\le1/(K_3)^2$ be some constant, define 
\be\label{adoq}
\RR^\cQ_T:=\{\lambda\ \vert |\lambda\log^2(\mu_0T)|<c \},
\ee 
then the perturbation series of $S_{\cQ}(\lambda)$ is convergent for $\lambda\in\RR^\cQ_T$.
This concludes the theorem.
\end{proof}

\begin{remark}\label{rmtad}
Obviously $\RR^{\cQ}_T\subset\RR^{\cC}_T$. This fact also set a constraint to the analytic domains for the biped Schwinger functions. Define the analytic domain for the two-point Schwinger functions by $\RR_T$, then we have 
\be\label{d2p}
\RR_T=\RR_T\cap\RR^{\cC}_T\cap \RR^{\cQ}_T\subseteq\RR^{\cQ}_T.
\ee 
Therefore, in order that the perturbation series of the $2p$-point, $p\ge1$, Schwinger functions to be convergent, the coupling constant $\lambda$ should be bounded by 
\be
|\lambda|<c/{(j_{max}+j_0)^2}=c/{\log^2 (\mu_0T)}.
\ee
\end{remark}
\section{The $2$-point Functions}
In this section we study the connected $2$-point Schwinger functions and the self-energy functions. While the perturbation series for the former are labeled by connected graphs, the ones for the latter are labeled by the one-particle irreducible graphs (1PI for short), which are the graphs that can't be disconnected by deleting any internal line. 
Let $S^c_{2,\b}(y,z,\l)$ be a connected $2$-point Schwinger function, in which $y$ and $z$ are the coordinates of the two external fields. Using the BKAR tree formula (\cite{RW1}, Theorem $4.1$) and organizing the perturbation terms according to the GN trees, we have:
\bea\label{consch}
S^c_{2,\b}(y,z,\l)&=&\sum_{n=0}^\infty\frac{\lambda^{n+2}}{n!}\int_{({\Lambda_{\beta}})^n} d^3x_1\cdots d^3x_n\sum_{\cG^{r_{max}}}\sum_{\cG^{r_{max}}_\cB}\sum_{{\cal EB}}\sum_{\{\sigma\}}\sum_{\cT, \tau}\Big(\prod_{\ell\in\cT}\int_{0}^1 dw_\ell\Big)\nn\\
&&\quad\quad\cdot \Big[\prod_{\ell\in\cT}C(f_\ell,g_\ell)\Big]\cdot\det\Big(C(f,g,\{w_\ell\})\Big)_{left},
\eea
where $\cG^{r_{max}}_\cB$ is a biped GN tree with root index $r_{max}$ (cf. \cite{RW1}, Definition $5.1$) and $\cT$ is a spanning tree in the root graph of $\cG^{r_{max}}_\cB$. The self-energy $\Sigma(y,z)$ are obtained by Legendre transform on the generating functional for $S^c_{2,\b}$ (cf. \eqref{self1}). In terms of Feynman graphs this corresponds to replacing the connected graphs labeling the connected functions by the 1PI graphs (also called the two-connected graphs \cite{graph}). Let $\{\Gamma\}$ be the set of 1PI graphs over the $n+2$ vertices, the self-energy is (formally) defined by:
\bea\label{selfeng}
&&\Sigma(y,z,\lambda)=\sum_{n=0}^\infty\frac{\lambda^{n+2}}{n!}\int_{({\Lambda_{\beta}})^n} d^3x_1\cdots d^3x_n\sum_{\cG^{r_{max}}}\sum_{\cG^{r_{max}}_\cB}\sum_{{\cal EB}}\sum_{\{\sigma\}, \tau}\sum_{\{\cT\}}\sum_{\{\Gamma\}}\\
&&\quad \Big(\prod_{\ell\in\cT}\int_{0}^1 dw_\ell\Big)\cdot \Big[\prod_{\ell\in\cT}C(f_\ell,g_\ell)\Big]\Big[\prod_{\ell\in\Gamma\setminus\cT}C(f_\ell,g_\ell)\Big]
\cdot\det\Big(C(f,g,\{w_\ell\})\Big)_{left,\Gamma},\nn
\eea
in which the summation is not over {\it all of 1PI graphs} but the ones generated in the multi-arch expansion, which will be introduced shortly. 
\begin{remark}
Remark that the bounds for the $2$-point Schwinger function and the self-energy function require the knowledge of the full Gallavotti-Nicol\`o tree structure. Since the contributions from the quadruped graphs and convergent graphs are convergent for $\l\in\RR_T$, we can safely omit the contributions from these of GN trees, but use only the fact that the analytic domain for the biped Schwinger functions can't be larger than $\RR_T$. 
\end{remark}
\subsection{Localization of the two-point Schwinger function}
The construction of the 2-point Schwinger functions and the self-energy requires renormalization. In the first step we introduce the localization operator for the two-point Schwinger function, which is naturally defined in the momentum space.
Let $\hat S_2(k)$ a two-point function with external momentum $k$.
Suppose that the internal momentum of the lowest scale belongs to the sector $(j_r, s_{+,j_r}, s_{-,j_r})$ while $k$ belongs to the sector with scale index $j_e$ and sector indices $(s_{+,j_e}, s_{-,j_e})$. The localization operator is defined as:
\be
\tau\hat S_2(k,\l)=\sum_{j=1}^\infty\sum_{\sigma=(s_+,s_-)}\chi_j(k_0^2+e^2(\bk))\cdot v_{s_+}(q_+^2)
\cdot v_{s_-}(q_-^2)\cdot\hat S_2 (2\pi T, \bk_F,\l),
\ee
in which $\bk_F=P_F(\bk)\in\cF_0$. Consider the integral
\bea\label{rn2pt0}
I=\int_{\cD_{\beta,L}\times\cD_{\beta,L}} dp dk\ \hat S_{2}
(k,\l)\hat C(p)\hat R(k,p,P_e),
\eea
in which $\hat C(p)=\sum_j\sum_{\sigma=(s_+,s_-)}\hat C_{j,\sigma}(p)$ is the sectorized free propagator, $\hat R(k,p,P_e)=\bar R(k,P_e)\delta(k-p)$ and $P_e$ are the external momentum. Define the localization operator $\tau$ on $I$ by:
\be
\tau I=\int dp dq\ \hat S_{2}
(2\pi T,\bk_F,\l)\hat C(q) \hat R(p,k,P_e),
\ee
and the remainder term is defined as $\hat R I=(1-\tau)I$. In order to establish the non-perturbative bound, it is also important to consider the localization in the direct space. The direct space representation of $I$ is given by:
\bea\label{rn2pt1}
\tilde I=\int dy dz\ S_{2} (x,y,\l)\ C(y,z)R(z,x,P_e),
\eea
which is indeed independent of $x$, due to translational invariance. Then the operators $\tau$ and $\hat R:=(1-\tau)$ induce the actions $\tau^*$ and $\hat R^*:=(1-\tau^*)$ in the direct space. The localized term is
\be
\tau^*\tilde I=\int dy dz\ S_{2}(x,y,\l)[e^{ik^0_F(x_0-y_0)+i\kk_F\cdot (\xx-\yy)} C(x,z)]R(z,x,P_e),
\ee
in which $k^0_F=2\pi T$. Comparing with \eqref{rn2pt1} we find that the localization operator moves the starting point $y$ of the free propagator to the localization point $x$, with the compensation of a phase factor:
\be
\tau^*C(y,z)=e^{i2\pi T(x_0-y_0)+i\kk_F\cdot (\xx-\yy)} C(x,z).
\ee
The remainder term is:
\bea
\hat R^* I=\int dy dz\ S_{2}(x,y)[C(y,z)-e^{i2\pi T(x_0-y_0)+i\kk_F\cdot (\xx-\yy)} C(x,z)]R(z,x,P_e),
\eea
in which
\bea
&&C(y,z)-e^{i2\pi T(x_0-y_0)+i\kk_F\cdot (\xx-\yy)} C(x,z)\\
&&=\int_0^1 dt(y_0-x_0)\frac{\partial}{\partial x_0}C((ty_0+(1-t)x_0,\yy),z)\nn\\
&&+\frac12(y_+-x_+)(y_--x_-)\partial_{x_+}\partial_{x_-}
C((x_0,\xx),z)\nn\\
&&+\int_0^1 dt(1-t)(y_+-x_+)(y_--x_-)\partial_{y_+}\partial_{y_-}
C((x_0, t\yy+(1-t)\xx),z)\nn\\
&&+C((x_0,\xx),z)[1-e^{i2\pi T(x_0-y_0)+i\kk_F\cdot (\xx-\yy)}]\nn.
\eea
The last line means that there exists an additional propagator $ C(x,z)$ attached on the biped graph so that the new graph has three external lines and is not linearly divergent anymore. Now we consider the other terms. Suppose that the internal momentum of the two-point function $S_{2}(x,y,\l)$ of the lowest scale belongs to the sector $\Delta^{j_r}_{s_{j_r,+},s_{j_r,-}}$, the sector with scale index ${j_r}$ and sector indices $({{s_{j_r,+},s_{j_r,-}}})$, while the external momentum belongs
to the sector $\Delta^{j_e}_{s_{j_e,+},s_{j_e,-}}$,
then there exists a constant $K_1$, $K_2$ such that
\bea\label{rmdx}
&&|y_0-x_0|\le O(1)\gamma^{j_r},\  
|\partial_{x_0}C((x_0,\yy),z)|\le K_1\g^{-j_e}|C((x_0,\yy),z)|,\nn\\
&&|y_+-x_+|\cdot|y_--x_-|\le\g^{s_{j_r,+}+s_{j_r,-}},\ |\partial_{x_+}\partial_{x_-}C((x_0,\xx),z)|\le K_2\g^{-s_{j_e,+}-s_{j_{e,-}}},\nn
\eea
and we have $j_r\le j_e$, due to the Gallavotti-Nicol\`o tree structure. As will be proved in Section $7.2$, we can always choose the optimal internal propagators (rings propagators) such that $s_{j_e,+}+s_{j_{e,-}}\ge s_{j_r,+}+s_{j_{r,-}}$, so that we can gain a convergent factor
$\g^{-(j_e-j_r)}$ and $\g^{-[(s_{j_e,+}+s_{j_{e,-}})-(s_{j_r,+}+s_{j_{r,-}})]}$.
The localization for the self-energy can be easier performed in the momentum space. Let $\hat \Sigma(k_0,\bk,\l)_{s_+,s_-}$ be the Fourier transform of $\Sigma(x,\lambda)$ such that the internal momentum of the lowest scale belongs to the sector $\Delta^{j_r}_{s_{j_r,+},s_{j_r,-}}$ and the external momenta are constrained to the sector $\Delta^{j_e}_{s_{j_e,+},s_{j_e,-}}$. The localization operator is defined as:
\be
\tau\hat\Sigma(k_0,\bk,\l)_{s_+,s_-}=\sum_{j=1}^{j_{max}}\sum_{\sigma=(s_+,s_-)}\chi_j(k_0^2+e_0^2(\bk))\cdot v_{s_+}(q_+^2)
\cdot v_{s_-}(q_-^2)\cdot\hat\Sigma (2\pi T, \bk_F,\l)_{s_+,s_-},\label{locsef}
\ee
and the remainder term is:
\bea\label{rmd11}
&&\hat R\hat \Sigma(k_0,\bk,\l)_{s_+,s_-}:=(1-\tau)\hat \Sigma(k_0,\bk,\l)_{s_+,s_-}\\
&=&\int_0^1 dt (k_0-k_F^0)\frac{\partial}{\partial k_0(t)}\hat \Sigma(k_F^0+t(k_0-k_F^0,\l),\bk)_{s_+,s_-}\nn\\
&+&\int_0^1 dt(1-t)(p_+-k_{F,+})(p_--k_{F,-})\frac{\partial^2}{\partial q_+\partial q_-}\hat \Sigma(k_F^0,\bk_F+t\bq,\l)_{s_+,s_-}\nn\ ,
\eea
where $k_0(t)=k_F^0+t(k_0-k_F^0)$ and $k_\pm(t)=k_{F,\pm}+tq_\pm$. 
We have $|k_0-k_F^0|\sim \gamma^{-j_e}$, $\Vert{\partial k_0}\hat \Sigma(k_F^0,\bk)_{s_+,s_-}\Vert_{L^\infty}\sim\gamma^{j_r}\Vert\hat \Sigma(k_F^0,\bk)_{s_+,s_-}\Vert_{L^\infty}$. So that
\be\label{rmd12}
\Vert(k_0-k_F^0)\frac{\partial}{\partial k_0}\hat \Sigma(k_F^0,\bk_F+\bq,\l)_{s_+,s_-}\Vert_{L^\infty}\le K_1\gamma^{-(j_e-j_r)}\Vert\hat \Sigma(k_F^0,\bp_F+\bq,\l)_{s_+,s_-}\Vert_{L^\infty},
\ee
and
\bea\label{rmd13}
&&\Vert(q_+q_-\frac{\partial^2}{\partial q_+\partial q_-}\hat \Sigma(k_F^0,\bk_F+\bq,\l)_{s_+,s_-}\Vert_{L^\infty}\\
&&\quad\quad\quad\le
K_2\g^{-[(s_{j_e,+}+s_{j_{e,-}})-(s_{j_r,+}+s_{j_{r,-}})]}\Vert\Sigma(k_F^0,\bk_F+\bq,\l)_{s_+,s_-}\Vert_{L^\infty}\nn,
\eea
for some positive constants $K_1$ and $K_2$. In conclusion, we gain a convergent factor $$\g^{-[(s_{j_e,+}+s_{j_{e,-}})-(s_{j_r,+}+s_{j_{r,-}})]}.$$
\subsection{The renormalization}
The renormalization analysis is to be performed in multi-steps according to the GN trees. At any scale $r$, we shall move the counter-terms from the interaction to the covariance, 
so that the tadpoles as well as the self-energy at that scale can 
be compensated by the counter-terms and the renormalized band function $e(k,\mu,\l)$ remain fixed.
\subsubsection{Renormalization of the bare chemical potential}
In this part we consider the renormalization of the bare chemical potential $\mu_{bare}=\mu+\delta\mu$, realized by the compensation of the tadpoles $T(\l)$ with the counter-term $\delta\mu$. Define
\be
\delta\mu(\lambda)=\delta\mu^{\le r_{max}}(\lambda)=\sum_{r=0}^{r_{max}} \delta\mu^r(\lambda),\ T(\lambda)=T^{\le r_{max}}(\lambda)= \sum_{r=0}^{r_{max}} T^r(\lambda),
\ee 
in which $T^r(\lambda)\in\RRR$ is the sliced tadpole with the {\it internal momentum} constrained to the $r$-th shell. We have:
\begin{lemma}\label{tadmain1}
Let $ T(\l)$ be the amplitude of a tadpole. There exist two positive constants $c_1'<c_2'$, such that:
\be
c_1'|\lambda|\le\vert T\vert\le c_2'|\lambda|.
\ee
\end{lemma}
\begin{proof}
It is more conveniently to express the amplitude as $T=\sum_{j=0}^{j_{max}}T^j$. For any scale index $0\le j\le j_{max}$, we have:
\be
\vert T^j\vert=|\lambda|\sum_{\s=(s_+,s_-)}\ \Big|\int dk_0 dq_+dq_- C_{j,\sigma}(k_0,k_{F,+}+q_+,k_{F,-}+q_-)\ \Big|= c_1 |\lambda| \sum_{(s_+,s_-)}\g^{-s_+-s_-}.\nn
\ee
When $j\le j_0$, we have $0\le s_\pm\le j_0$, hence summing over $s_\pm$ is bounded by $c_1|\l|$ for some positive constant $c_1$. When $j>j_0$, we have $(j-j_0)/2\le s_+, s_-\le j$. Using the constraint $s_++s_-\ge 3j/2-j_0/2$, we have
\bea
&&\sum_{(s_+,s_-)}\g^{-s_+-s_-}=\Big(\sum_{s_+=(j-j_0)/2}^{j}\g^{-s_+}\Big)\ \Big(\sum_{s_-=3j/2-j_0/2-s_+}^{j} \g^{-s_-}\Big)\nn\\
&&\le \frac{j+j_0}{2}\g^{-3j/2+j_0/2}.
\eea
Hence there exist two positive constant $c'_2>c'_1>c_1$, which are independent of $j$, such that
\bea\label{bdtj}
c'_1|\lambda|(j+j_0)\g^{-3j/2+j_0/2}\le\vert T^j\vert\le c'_2|\lambda|(j+j_0)\g^{-3j/2+j_0/2}.
\eea
Since $|T|=\sum_{j}|T^j|$, the conclusion follows by summing over the scale index $j$.
\end{proof}

At each scale $r$, the chemical potential counter-term $\delta\mu^r$ is compensated with the tadpole term $T^r$ the renormalized chemical potential $\mu$ remain fixed. By locality, the compensations are $exact$. In order that Equation \eqref{rncd3} can be valid, we have:
\be\label{rnc}
T^r+\delta\mu^r(\lambda)=0,\  {\rm for}\ r=0,\cdots, r_{max},
\ee
which implies that:
\be\delta\mu^{\le r}(\lambda)+\delta T^{\le r}=0,\ {\rm for}\ r=0,\cdots, r_{max}.\ee
These cancellations are between a pair of GN trees. Let $F_{2,n}=F'_{2,n-1}\vert T^r\vert$ be the amplitude of a graph with $n$ vertices which contains a tadpole $T^r$. Let $F'_{2,n}=F'_{2,n-1}\delta\mu^r$ be the amplitude of another graph which contain counter-term, which is located at the same position in the GN tree as the tadpole. Then we have $F_{2,n}+F'_{2,n}=0$. 
In this way we can fix the renormalized chemical potential at all scales.
\begin{theorem}\label{flowmu}
There exists a positive constant $K$ independent of the scale indices such that 
the tadpole counter-term can be bounded as follows:
\be\label{bdbare}
|\delta\mu(\lambda)|:=|\delta\mu^{\le r_{max}}(\lambda)|\le K|\lambda|,\ {\rm for}\ \lambda\in\RR_T.
\ee
\end{theorem}
We will prove this theorem in Section \ref{secflow}. Consequently, we can always replace the term $\delta\mu$ by some positive constant.

\subsubsection{Renormalization of $\hat\Sigma(k,\l)$} 
Now we consider the renormalization of the non-local part of the self-energy. Since the cancellation between tadpoles and counter-terms are exact, we assume that the self-energy function $\hat\Sigma$ is tadpole free. Rewrite also the counter-term in the multi-scale representation as
\be
\hat\nu(\bk,\l)=\hat\nu^{\le r_{max}}(\bk,\l):=\sum_{r=0}^{r_{max}}\sum_{\sigma=(s_+,s_-)}\hat\nu_{s_+,s_-}^r(\bk,\l),
\ee
in which
\be
\hat\nu_{s_+,s_-}^r(\bk,\l)=\hat\nu(\bk,\l)\chi_j(k_0^2+e_0^2(\bk,\mu_0))\cdot v_{s_+}(q_+^2)
\cdot v_{s_-}(q_-^2).
\ee
The multi-scale representation for the self-energy function is:
\bea
\hat\Sigma(k,\hat\nu,\lambda )=\Sigma^{\le r_{max}}(k,\hat\nu^{\le r_{max}},\lambda ):=\sum_{r=0}^{r_{max}}\sum_{\sigma=(s_+,s_-)}\hat\Sigma_{s_+,s_-}^r(k,\hat\nu^{\le (r)},\lambda ),
\eea
in which
\be
\hat\Sigma_{s_+,s_-}^r(k,\hat\nu^{\le (r)},\lambda )=\hat\Sigma(k,\hat\nu^{\le (r)},\lambda )\chi_j(k_0^2+e_0^2(\bk,\mu_0))\cdot v_{s_+}(q_+^2)
\cdot v_{s_-}(q_-^2).
\ee
The localization for the self-energy is defined as (see also \eqref{locsef}): 
\be\tau \hat\Sigma_{s_+,s_-}^{r}\big[k,\hat\nu^{\le r},\lambda \big]=
\hat\Sigma_{s_+,s_-}^{r}\big[(2\pi T,P_F(\bk)_{s_+,s_-}),\hat\nu^{\le r},\lambda \big],
\ee
in which $P_F(\bk)_{s_+,s_-}\in\cF\cap \Delta^r_{s_+,s_-}$ is a projection of the vector $\bk\in\Delta^r_{s_+,s_-}$ on the Fermi surface. In each renormalization step we move the counter-term $\hat\nu^r$ from the interaction potential to the covariance, so that the localized self-energy term $\hat\Sigma^r$ can be compensated. The renormalization condition becomes:
\be\label{rs1}
\hat\Sigma_{{s_+,s_-}}^{r}\big[(2\pi T,P_F(\bk))_{s_+,s_-},\hat\nu^{\le (r)},\lambda \big]+
\hat\nu^{r}_{s_+,s_-}(P_F(\bk)_{s_+,s_-},\lambda)=0.
\ee
When the external momenta are not on the Fermi surface, the compensation is not to be exact, due to the non-locality of the proper self-energy $\hat\Sigma^r(k,\l)$ and the counter-term $\hat \nu^r(\bk,\l)$. Then the renormalization condition is defined as:
\be\label{rs2}
\hat\Sigma_{s_+,s_-}^r((k_0,\bk),\hat\nu^{\le (r)},\lambda )+\hat\nu^{r}_{s_+,s_-}(\bk,\l)=\hat R\hat\Sigma_{s_+,s_-}^{r}((k_0,\bk),\hat\nu^{\le (r+1)},\lambda ),
\ee
in which the remainder term $\hat R\hat\Sigma_{s_+,s_-}^{r}((k_0,\bk),\hat\nu^{\le (r+1)},\lambda )$ is bounded by $$\g^{-\delta^r}\Vert \hat\Sigma_{s_+,s_-}^{r+1}(k,\hat\nu^{\le (r+1)},\lambda )\Vert_{L^\infty},$$
in which 
\be\label{rmd14}
\g^{-\delta^r}=\max \{\g^{-(r_e-r_r)},  \g^{-[(s_{j_e,+}+s_{j_e,-})-(s_{j_r,+}+s_{j_r,-})]}\}<1.\ee 
From the renormalization conditions \eqref{rs1} and \eqref{rs2} we have:
\be\label{rmd15}
\Vert\hat\nu(\bk,\lambda)\Vert_{L^\infty}=\sup_{\bk}|\hat\nu(\bk,\lambda)|\le \sup_{\bk}\sum_{r=0}^{r_{max}}|(1+\hat R)\hat\Sigma^r(\bk,\lambda)|\le 2\sup_{\bk}\sum_{r=0}^{r_{max}}|\hat\Sigma^r(\bk,\lambda)|.
\ee
We have the following theorem concerning the bound for the counter-term:
\begin{theorem}\label{flownu}
There exists a positive constant $K$ independent of the scale indices such that 
\be\label{bdbarenu}
\Vert\nu^{\le r_{max}}(\bk,\lambda)\Vert_{L^\infty}\le K|\lambda|,\ \forall\ \lambda\in\RR_T.
\ee
\end{theorem}
By \eqref{rmd15}, in order to prove this theorem, it is enough to prove that there exists a positive constant $K$ such that $\Vert\hat\Sigma^{\le r_{max}}\Vert_{L^\infty}\le\frac{K|\lambda|}{2}$, which will be proved in Section \ref{multiarch}.


\section{Multi-arch Expansion for the Self-energy Function}\label{multiarch}
\begin{figure}[htp]
\centering
\includegraphics[width=.6\textwidth]{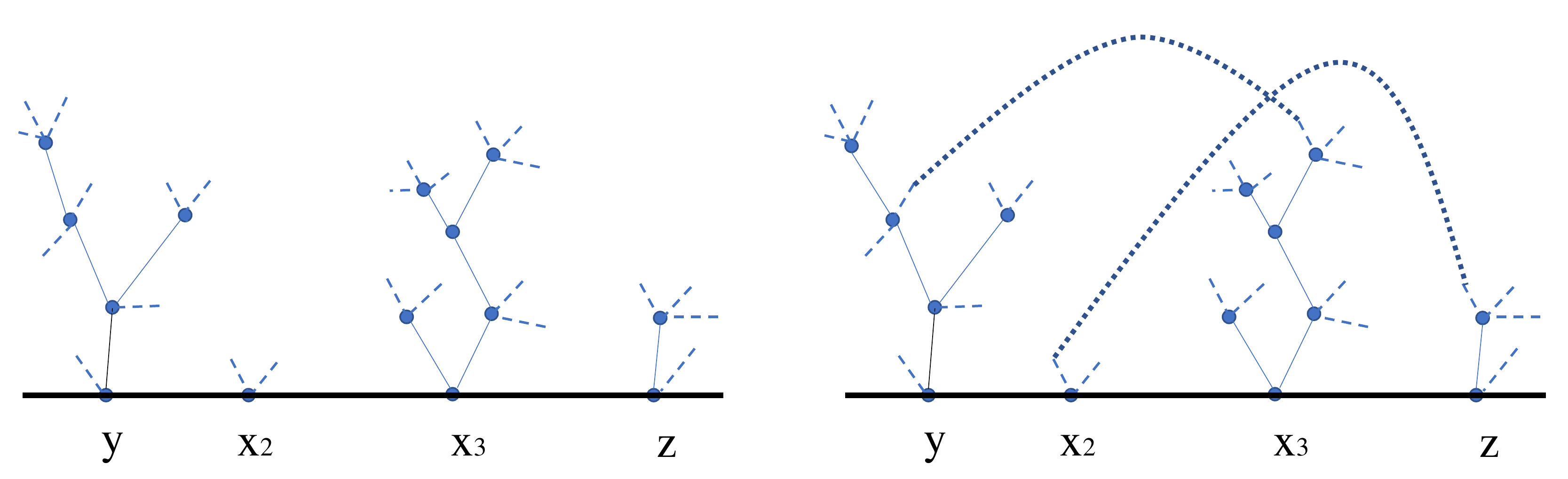}
\caption{\label{mtarch}
An illustration of the multi-arch expansion.}
\end{figure}

In this section we establish the upper bounds for the self-energy function and its derivatives w.r.t. the external momenta. The main tool is the multi-arch expansion for the determinant, which is a canonical way of generating the perturbation series of the self-energy labeled by the 1PI graphs between any two external vertices, say $y$ and $z$, without generating any divergent combinatorial factor. This expansion has been introduced in great detail in \cite{RWang1}, Section 7.1 or \cite{AMR1}, Section VI. So we only sketch the main idea of this expansion and ask the interested reader to consult \cite{RWang1} or \cite{AMR1} for more details. Let $\cT$ be a spanning tree connecting the $n+2$ vertices $\{y,z,x_1\cdots,x_n\}$. The integrand of the connected two-point function labeled by $\cT$ is
\be\label{main0}
F(\{C_\ell\}_\cT, \{C(f_i,g_j)\})=\Big[\prod_{\ell\in\cT}C_{\s(\ell)}(f(\ell),g(\ell))\Big]\ \det(\{C(f_i,g_j)\})_{left,\cT}.
\ee
Let $P(y,z,\cT)$ be the unique path in $\cT$ such that $y$ and $z$ are the two ends of the path. Suppose that there are $p+1$ vertices in the path $P(y,z,\cT)$, $p\le n+1$. Then we can label each vertex in the path with an integer, starting from the label $0$ for the vertex $y$ in an increasing order towards $p+1$, which is the label for the vertex $z$. Let $\mathfrak{B}_i$ be a branch in $\cT$ at the vertex $i$, $1\le i\le p+1$, which is defined as the
subtree in $\cT$ whose root is the vertex $i$. See Figure \ref{mtarch} for an illustration of a tree graph with $4$ branches rooted at the four vertices $y, x_1,x_2$ and $z$.
We fix two half lines, also called the external fields, each is attached to one end vertex. There are in total $4n$ fields associated to the $n$ vertices $\{x_1\cdots,x_n\}$. Since each tree line in $P(y,z,\cT)$ contracts $2$ fields, there are still $2(n+2)$ fields to be contracted from the determinant $\det_{left}$, which are called the {\it remaining fields}. The set of remaining fields is denoted by $\mathfrak{F}_{left}$. A packet $\mathfrak{F}_i$ is defined as the set of the remaining fields restricted to a branch $\mathfrak{B}_i$. By definition we have: $\mathfrak{F}_i\cap\mathfrak{F}_j=\emptyset$ for $i\neq j$, and $\mathfrak{F}_{left,\cT}=\mathfrak{F}_1\sqcup\cdots\sqcup\mathfrak{F}_p$, in which $\sqcup$ is the disjoint union. In each step of the multi-arch expansion we expand a loop line, also called {\it an arch}, from the determinant by contracting
an element of $\sqcup_{k=1}^{p_1}\mF_i$, $p_1\le p+1$, and an element of $\sqcup_{k=p_1+1}^{p}\mF_k$, through an explicit Taylor expansion with interpolating parameter (see \cite{RWang1}, Section 7.1, or \cite{AMR1}, Section VI). Suppose that $m$ arches have been generated to form a 1PI graph, the set of arches 
\bea
&&\Big\{ \ell_1=(f_1,g_1),\cdots,\ell_m=(f_m,g_m)\ \Big|\ f_1\in\mF_1, g_1\in\mF_{k_1};\ f_2\in\sqcup_{u=1}^{k_1}\mF_u, g_2\in\sqcup_{u=k_{1}+1}^{k_2}{\mF_u};\nn\\
&&\quad\quad\quad \cdots ;\ 
f_m\in\sqcup_{u=1}^{k_{m-1}}\mF_u,\ g_m\in\mF_{k_m}=\mF_{p};\  k_1\le\cdots\le k_m;\ m\le p\ \Big\}.
\eea
is called an {\it m-arches system}.
Given an $m$-arch system, one can choose a {\it minimal} $m$-arch subsystem system (\cite{AMR1}, Section VI.1) which contains minimal number of loop lines. So we always assume that the $m$-arch system is minimal. In order to obtain the optimal bounds for the self-energy function, a second multi-arch expansion has to be performed on top of the first one, and the resulting graphs are {\it two-particle irreducible} (2PI for short) between the two external vertices, which means that one can't disconnect this graph by deleting any two internal lines. 
We have:
\begin{lemma}
The amplitude for the self-energy reads
\bea
&&\Sigma (y,z,\l)=\sum_{n=2}^\infty \frac{\lambda^{n}}{n!} \int_{(\Lambda_\b)^n} d^3x_1 ... d^3x_n 
\sum_{\{ \t  \}} \sum_{\substack{\text{biped structures}\\ \mathcal{B}}
}\sum_{ \mathcal{EB}}\nn\\
&&\quad\quad\sum_{\cG^{r_{max}}_\cB} \sum_\mathcal{T} \sum_{\{\sigma \}}\sum_{\substack{ m-{\rm arch\ systems}\\ \bigl( (f_1,g_1), ... (f_m,g_m) \bigr)}}
\sum_{\substack{ m'-{\rm arch\ systems}\\ \bigl( (f'_1,g'_1), ... (f'_m,g'_m) \bigr)}}
\nn\\
&&\quad\quad\left( \prod_{\ell \in \mathcal{T}} \int_0^1 dw_\ell \right) \left( \prod_{\ell \in \mathcal{T}} C_{\sigma(\ell)} (f_\ell,g_\ell)\right)\left( \prod_{u=1}^m \int_0^1 ds_u \right)\left( \prod_{u'=1}^{m' }\int_0^1 ds'_{u'} \right)\nn \\
&&\quad\quad\left( \prod_{u=1}^m C(f_u,g_u) (s_1,...,s_{u-1})\right) \left( \prod_{u' = 1}^{m'} C({f'}_{u'},{g'}_{u'}) (s'_1,...,s'_{u'-1})\right)\nn\\
&&\quad\quad\frac{\partial^{m+m'} \det_{\text{left}, \mathcal{T}}}{\prod_{u=1}^m \partial C(f_u,g_u)\prod_{u'=1}^{m'} \partial C(f'_{u'},g'_{u'}) }\big( \{ w_\ell\}, \{ s_u\} ,  \{ s'_{u'} \}  \big),
\eea
where we have summed over all the first multi-arch systems with $m$ loop lines and
the second multi-arch systems with $m'$ loop lines. The underlying graphs are 2PI.
\end{lemma}



\begin{figure}[htp]
\centering
\includegraphics[width=.42\textwidth]{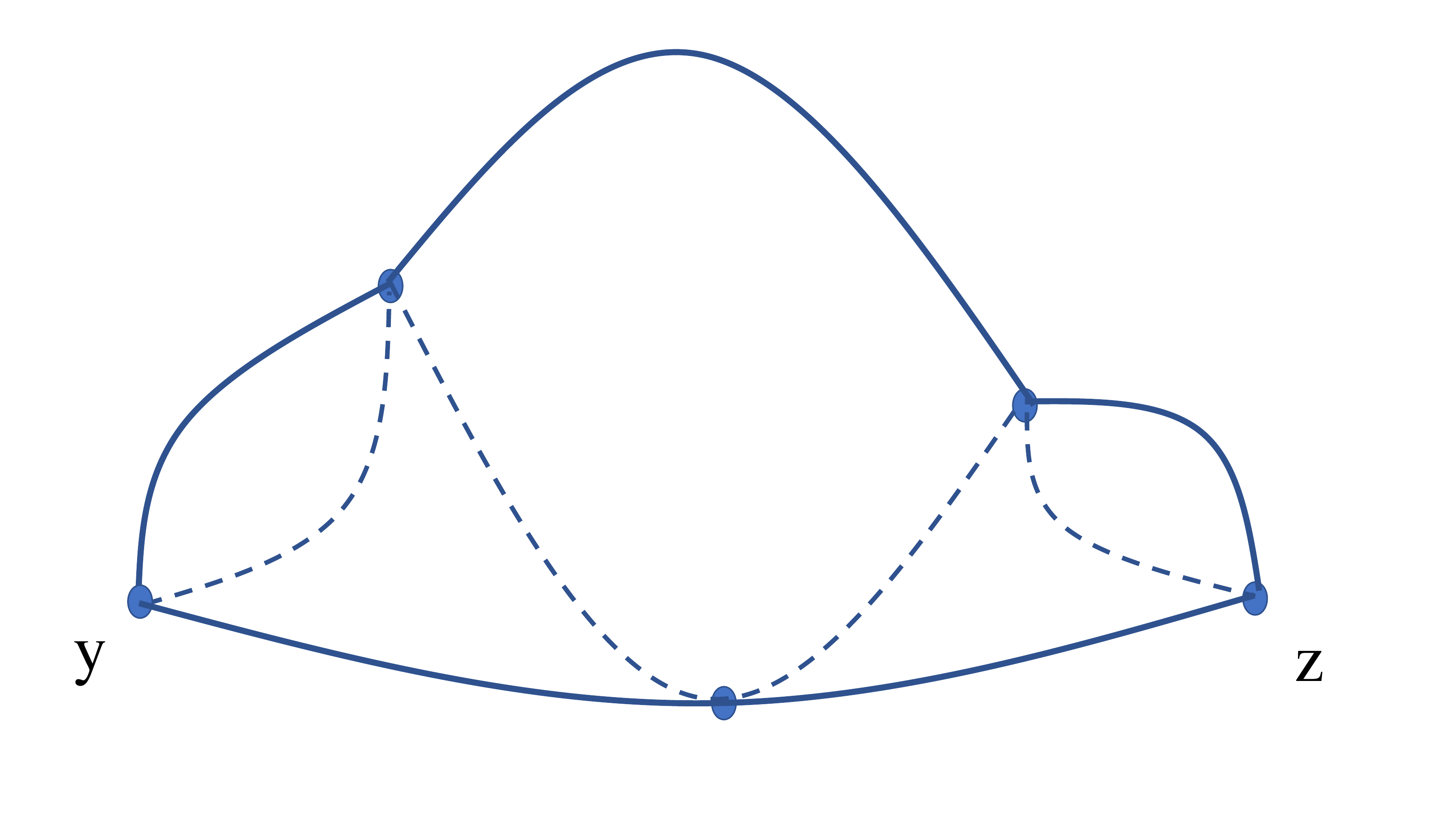}
\caption{\label{rin}
A ring structure in a $2$PI graph which connects the two end vertices $y$ and $z$. The thick lines are the ring propagators and the dash lines are the third path in the $2$PI graph.}
\end{figure}

Let $G=\cT\cup\cL$ be a 2PI graph generated by the two-level multi-arch expansions. Menger's theorems (\cite{graph}, Section 3.3) ensure that $G$ has three line-disjoint independent paths and two internally vertex-disjoint paths joining the two external vertices of $G$. In order to obtain the optimal bounds for the self-energy, we need to choose the optimal integration paths in the graph from which we can obtain the best convergent scaling factors (see \cite{AMR1}, Section VI). The optimal paths are called the ring structure. See Figure \ref{rin} for an illustration.
\begin{definition}
A ring $R$ is a set of two paths $P_{R,1}$, $P_{R,2}$ in $\cL\cup\cT$ connecting the two external vertices $y$ and $z$ and satisfies the following conditions: (a) the two paths in $R$ don't have any intersection on the paths or on the vertices, except on $y$ and $z$. (b), let $b$ be any node in the biped tree $\cG^{r_{max}}_\cB$, then at least two external fields of $b$ are not contained in $R$.
\end{definition}
Let $P_{R,1}$ and $P_{R,1}$ be the two path in the ring and $k$ be the labeling of propagators in the two paths. Let $r_T$ be the first scale at which $y$ and $z$ fall into a common connected component in the GN tree. Define
\be
r_R=\min_{j=1,2}\max_{k\in P_{R,j}}r(k)\nn
\ee  
as the first scale at which the ring connects $y$ and $z$, called the the scale index for the ring propagators. Obviously we have: $r_T\le r_R\le r_{max}$. In the same way, we can define the sectors indices for the ring propagators: 
\be
s_{+,R}:=\min_{j=1,2}s_{+,R,j}:=\min\max_{k\in P_{R,j}}s_+(k),\  {\rm and}\ \ s_{-,R}:=\min_{j=1,2}s_{+,R,j}:=\min\max_{k\in P_{R,j}}s_-(k),
\ee
which are greater than the sector indices of the tree propagators. The corresponding sectors are called the {\it ring} sectors. 
We can also define the scale index $j_{R,T}$ for the ring propagators, as follows. Let $j_\cT=\max_{k\in P(y,z,\cT)}j_k$, in which  $P(y,z,\cT)$ is the unique path in $\cT$ connecting $y$ and $z$. Define $j_R=\min_{j=1,2}\max_{k\in P_{R,j}}j(k)$, $j_{R,T}=\min\{j_R, j_\cT\}$, and $
r_{R,\cT}=(j_{R,T}+s_{+,R}+s_{-,R})/2$, which is the $r$-index for the ring sectors. With all these preparations, we can prove the following upper bound for the self-energy function:
\begin{theorem}\label{maina}
Let $\Sigma(y,z,\l)^{\le r}$ be the biped self-energy function with root scale index $r$, denote the corresponding Gallavotti-Nicol\`o tree by $\cG^r$. Let $\lambda\in\RR_T:=\{\l\vert |\lambda\log^2\mu_0T|\le K_1 \}$, in which $K_1$ is some positive constant independent of $\l$ and the scale index (cf. \eqref{adoq}) and let $\sigma_{{\cG^r}}$ be the set of sector indices that is compatible with the Gallavotti-Nicol\`o tree $\cG^r$. Then there exists a constant $K$, which is independent of the scale index, such that:
\be\label{x2pta}
|\Sigma(y,z,\l)^{\le r}|\le K\lambda^2 r^2\sup_{\sigma\in\sigma_{\cG^r}}\g^{-3r(\sigma)}e^{-c[d_{j,\s}(y,z)]^\a},
\ee
\be\label{x2pt}
|z^{+}-y^{+}||z^{-}-y^{-}||\Sigma(y,z,\l)^{\le r}|\le K\lambda^2 r^2\sup_{\sigma\in\sigma_{\cG^r}} \g^{-2r(\sigma)}e^{-c[d_{j,\s}(y,z)]^\a},
\ee
\be\label{x2ptc}
|y_0-z_0||\Sigma(y,z,\l)^{\le r}|\le K\lambda^2 r^2\sup_{\sigma\in\sigma_{\cG^r}} \g^{-2r(\sigma)}e^{-c[d_{j,\s}(y,z)]^\a},
\ee
where $[d_{j,\s}(y,z)]^\a$ (cf. \eqref{dist0}) characterize the decaying behaviors of the propagator in position spaces. The upper bounds presented in \eqref{x2pta}, \eqref{x2pt} and \eqref{x2ptc} are optimal.
\end{theorem}

Before proving this theorem, we consider the following lemma concerning sector counting for the biped graphs. Instead of using the scale index $r$, it is more convenient to use the scale index $j$ and $j_0$.
\begin{lemma}[Sector counting lemma for bipeds]\label{secbi}
Let $b_r$, $r=(j+s_++s_-)/2$, be a 2PI biped with root scale index $r\in[0, r_{max}]$ and contains $n+2$ vertices. There exists a positive constant $C_1$, independent of the scale index $j$ and $j_0$, such that the summation over all the sector indices is bounded by $C_1^{n+2}(j+j_0)^{2n+1}$. 
\end{lemma}
\begin{proof}
The proof of this lemma is very similar to the one of \cite{RWang1}, Lemma 7.4, except that in the current case summation over the sector indices for each root field is bounded by $C_1(j+j_0)$, for some positive constant $C_1$ independent of $j$ and $j_0$ (cf. Lemma \ref{sec1}). The interested readers are invited to consult \cite{RWang1} for more details.
\end{proof}

\begin{proof}[Proof of Theorem \ref{maina}]
The proof of this theorem is almost identical to \cite{AMR1}, Section VIII, except that we have different scaling behaviors due to the different definition of sectors. We only sketch the main idea for the proof and ask the interested readers to consult \cite{AMR1} for more details.
After integrating out the parameters $\{w_\ell\},\{s_\ell\}$ for the tree expansions and multi-arch expansions, we can write the amplitude of the self-energy as:
\bea
|\Sigma^{\le r}(y,z,\l)|\le\sum_{n=2}^\infty\frac{(K\l)^{n}}{n!}\sum_{\cG_{\cB}}
\sum_{ \mathcal{EB}}  \sum_{ \mathcal{T} } \sum_{R}\sum_{\{ \sigma \}}\prod_p\chi_p(\sigma)\cI_{1,n}(y,z)\cI_{2,n}(y,z,x_{p,\pm}),
\eea
in which 
\be
\cI_{2,n}(y,z,x_{p,\pm})=\int\prod_{v\in R,v\neq y,z} dx_{v,0}\prod_{v\notin R}d^3x_v
\prod_{f\notin R}\g^{-r_f/2-l_f/4-(j_f-j_0)/8}\prod_{p\in \cL}e^{-\frac{c}{2}[d_{j,\s(p)}(y,z)]^\a}
\ee
is the factor in which we keep the position $y$ and $z$ and the spatial positions of the ring vertices $x_{p,\pm}$ fixed and integrate out all the remaining positions. A fraction (one half) of the decaying factor from every loop line in $\cL$ and from the remaining determinant has been also put here to compensate possible divergence from the integrations. The factor
\be
\cI_{1,n}=\int\prod_{i=1}^p dx_{i,+}dx_{i,-}\prod_{k\in R}\g^{-[r_k+l_k/2+(j_k-j_0)/4]}\prod_{p\in\cL}e^{-\frac{c}{2}[d_{j,\s(p)}(y,z)]^\a},
\ee
in which $x_i$ are the internal vertices other than $y$ and $z$ in the ring, contains the remaining terms and integrations, and can be derived with the same method as \cite{AMR1}, Formula (VIII.92). Following the same analysis as \cite{AMR1} Lemma VIII.1 and VIII.2, we can prove that:
\be
\vert\cI_{2,n}\vert\le K_1^n \g^{-j_{\cT}}
\ee
and
\be
\vert\cI_{1,n}\vert\le K_2^p \g^{-s_{+,R,1}-s_{+,R,2}-s_{-,R,1}-s_{-,R,2}}e^{-\frac{c}{4}[d_{j,\s(p)}(y,z)]^\a},
\ee

Combining these two factors and summing over all the tree structure $\cT$, the ring structure $R$ as well as the GN trees, we have
\be
|\Sigma^{\le r}(y,z,\l)|\le \sum_n  \sum_{r'=0}^r\sum_{\{\sigma\}} C^n\lambda^{n} \g^{-j_{\cT}}\g^{-s_{+,R,1}-s_{+,R,2}-s_{-,R,1}-s_{-,R,2}}e^{-c'[d_{j,\s(p)}(y,z)]^\a},
\ee
for some positive constant $C$. Now we consider summation over the sector indices. Let $n=N+2$ and using Lemma \ref{secbi}, we obtain
\bea\label{finalsum}
&&\sum_{N=0}^{\infty}  \sum_{r'=0}^r\sum_{\{\sigma\}} C^{N+2}\lambda^{N+2} \le
\sum_{N=0}^{\infty}  \sum_{r'=0}^rC^{N+2}C_1^{N}\lambda^{N+2} r'^{2N+1}\nn\\
&&\quad\le \sum_{N=0}^{\infty} C_3^N[\lambda (j+j_0)^2]^N \l^2(j+j_0)^2,\label{ineq1}\\
&&\quad \le
\sum_{N=0}^{\infty} C_2^N(\lambda r^2)^N \l^2r^2,\label{ineq2}
\eea
for some positive constants $C_2$ and $C_3$ depending on $C$ and $C_1$. To obtain \eqref{ineq2} from
\eqref{ineq1} we have used the fact that there exists two positive constants $c_1$ and $c_2$ such that $c_1r\le j+j_0\le c_2r$ for $j_0$ fixed.
Both inequalities will be used depending the context.
Since $|\lambda(j+j_0)^2|\le c$ for $\lambda\in\RR_T$, summation over $N$
in \eqref{finalsum} is convergent provide that $c\cdot C_2<1$. This inequality can always be true for $c$ small enough.
So we obtain:
\bea\label{taylorm1}
|\Sigma^{\le r}(y,z,\l)|&\le& \sum_{N=0}^{\infty} C_2^N(\lambda r^2)^N \l^2r^2\sup_{\sigma\in\sigma_\cG}\g^{-3r(\sigma)}e^{-c'[d_{j,\s}(y,z)]^\a}\\
&\le&
 K_3\lambda^2 r^2\sup_{\sigma\in\sigma^r_\cG}\g^{-3r(\sigma)}e^{-c'[d_{j,\s}(y,z)]^\a},\nn
\eea
for certain positive constants $K_3$ and $c'$. By choosing the ring structure, the convergent factors obtained from $\cI_{1,n}$ and $\cI_{2,n}$ are optimal, so that the upper bounds we obtained for the self-energy are optimal. Following the same analysis for the derivatives of the self-energy, (see also \cite{AMR1}, pages 437-442) we can prove Formula \eqref{x2pt} and \eqref{x2ptc}. Hence we conclude Theorem \ref{maina}.
\end{proof}

We can also formulate Theorem \ref{maina} in the Fourier space, we have:
\begin{theorem}[Bounds for the self-energy in the momentum space.]\label{mainb}
Let $\hat\Si^{r}(\lambda,q)$ be the self-energy for a biped of scale index $r$ in the momentum space representation and $\lambda\in\RR_T$. There exists a positive constant $K$, which depends on the model but is independent of the scale index, such that:
\be \sup_q|\hat\Si^{r}(\lambda,q)|\le K\l^2 r\g^{-r},\label{spa}
\ee
\be \sup_q\vert \frac{\partial}{\partial q_\mu } \hat\Si^{ r} (\lambda,q) \vert\le K\lambda^2 r,\label{spb} 
\ee
\be\label{spc} \sup_q| \frac{\partial^2}{\partial q_\mu \partial q_\nu}  \hat\Si^{r} (\lambda,q) | \le K\lambda^2 r \g^{r},\ \mu,\nu\in\{+,-\}.
\ee
\end{theorem}
\begin{remark}\label{rmkb}
This theorem states that the self energy is uniformly $C^1$ in the external momentum for $|\lambda|<c/|\log \mu_0T|^2$ for some positive constant $c_1<1$, which is smaller than the one required by Salmhofer's criterion, which is $|\lambda|<c_1/|\log T|$. What's more, for $r=r_{max}$, with $\g^{r_{max}}\sim \frac1T$, then there exists some positive constant $C$, which is independent of the temperature, such that
\be\sup_q|\frac{\partial^2}{\partial q_\mu \partial q_\nu}  \hat\Si^{r} (\lambda,q) |\le\frac{C\l^2}{T},\ee 
which is not uniformly in $T$. Since the upper bound we obtained is optimal, this suggests that Salmhofer's criterion is violated and the ground state is not a Fermi liquid.
\end{remark}

%
%
\begin{proof}[Proof of Theorem \ref{mainb}]
Remark that the expressions for the self-energy function in \eqref{x2pta} and \eqref{spa} are simply Fourier dual to each other, so we just need to prove that
formula \eqref{spa}-\eqref{spc} can be derived from \eqref{x2pta}.
Let $\Sigma(y,z,\l)$ be the Fourier transform of $\hat\Sigma(k,\l)$. The left hand side of \eqref{x2pta} reads
\be
\Vert\Sigma^r(y,z,\l)\Vert_{L^\infty}=\Vert\sum_{k_0}\int dq\ \hat\Sigma^r(k_0,\bq+P_F\kk,\l) e^{ik(y-z)}\Vert_{L^\infty}=C \gamma^{-2r}\sup_q|\hat\Sigma^r(k_0,\bq+P_F\kk,\l)|,\nn
\ee
for some positive constant $C$ that is independent of $q$. Since 
\be
|\Sigma^r(y,z,\l)|\le C\lambda^2 r \g^{-3r},
\ee
we have 
\be
\sup_q|\hat\Sigma^r(k_0,\bq+P_F\kk,\l)|\le C\lambda^2 r \g^{-r}.
\ee
So we have proved \eqref{spa}. Since the momentum $q$ is bounded by $\g^{-r}$ at scale index $r$, the first order differentiation w.r.t. $q$ for the r.h.s. term of \eqref{spa} gives the bound in \eqref{spb} and the second order differentiation gives the bound in \eqref{spc}. 
The fact that these bonds are optimal also follow from the fact that the bounds in \eqref{x2pta}-\eqref{x2ptc} are optimal.
\end{proof}
As a corollary of Theorem \ref{mainb}, we have the following result:
\begin{theorem}\label{mainc}
There exists a constant $C$ which may depends on the model but is independent of the scale index $r$ and $\l$, such that the counter-term $\nu^{\le r}(\bk,\l)$ satisfies the following bound:
\be\label{cte1}
\sup_{\bk}|\nu^{\le r}(\bk,\l)|\le C\l^2 r\g^{-r}.
\ee
\end{theorem}
\begin{proof}
By renormalization conditions, we have
$\nu^{\le r}(\bk,\l):=-\tau\Sigma^r(k_0,P_F\bk)$, hence \be\sup_\bk|\nu^{\le r}(\bk,\l)|\le \sup_k|\hat\Sigma^r(k_0,\bk)|.\ee
Then we can prove this theorem using the bounds for the self-energy (cf. Equation \eqref{spa}).
\end{proof}


We have the following theorem concerning the $2$-point Schwinger function:
\begin{theorem}\label{maine}
Let $\hat S_2(k,\lambda)$ be the two-point Schwinger function. There exists a positive constant $K$ which is independent of $T$ and $\lambda$, such that for any $\lambda\in\RR_T$, we have
\be
\hat S_2(k,\lambda)=\hat C(k)[1+\hat R(\lambda, k)],
\ee
in which $\Vert \hat R(\lambda, k)\Vert_{L^\infty}\le K|\lambda|$ for some positive constant $K$ which is independent of $k$ and $\l$.
\end{theorem}
\begin{proof}
The proof is identical to the proof of Theorem 7.4 in \cite{RWang1} and we don't repeat it here.
\end{proof}


\subsection{Proof of Theorem \ref{conj2}}
With all these preparations, we are ready to prove Theorem \ref{conj2}.
\begin{proof}\label{prfconj}
In order to prove this Theorem, it is enough to verify item (2) of the renormalization conditions (cf. \eqref{rncd2}) at any 
slice $r\le r_{max}$ and prove that the counter-term $\nu(\bk,\l)$ is $C^{1+\epsilon}$ in $\bk$. First of all, by the multi-slice renormalization condition \eqref{rs1} and using \eqref{ineq1}, the ratio in \eqref{rncd2} at any slice $r$ is bounded by
\bea
&&\Vert\big[\hat\nu^{r}_{s_+,s_-}(\bk,\l)+\hat\Sigma^{r}_{s_+,s_-}(k,\l)\big]\hat C_r(k)\Vert_{L^\infty}\le K\cdot
\frac{\Vert\hat\Sigma^{r+1}_{s_+,s_-}(k,\l)\Vert_{L^\infty}}{\g^{-r}}\nn\\
&&\quad\le K_1\l^2\frac{(j+j_0)^2\g^{-r-1}}{\g^{-r}}=\g^{-1}\l^2 K'(j+j_0)^2
\le\l^2 K_2\log^2 \mu_0T,
\eea
in which $K$, $K_1$ and $K_2$ are some positive constants and we have used the relation  $\gamma^{j_{max}}\sim\frac1T$ and $\gamma^{j_0}\sim\frac{1}{\mu_0}$. 
Then we have
\be
\Vert\big[\hat\nu^{r}_{s_+,s_-}(\bk,\l)+\hat\Sigma^{r}_{s_+,s_-}(k,\l)\big]\hat C_r(k)\Vert_{L^\infty}
\le K_3|\lambda|,
\ee
which is bounded for $\l\in\RR_T$. By construction, the counter-term $\hat\nu^{r}_{s_+,s_-}(\bk)$ has the same regularity as the self-energy function $\Sigma^{r}_{s_+,s_-}(\bk)$. By Theorem \ref{mainb} we know that
the self-energy is $C^{1}$ (cf. Formula \eqref{spb}) but not $C^2$ (cf. Formula \eqref{spc}) w.r.t. the external momentum, so is the counter-term. Thus we proved Theorem \ref{conj2}.

\end{proof}

\subsection{Proof of Theorem \ref{flowmu}.}\label{secflow}
\begin{figure}[htp]
\centering
\includegraphics[width=.4\textwidth]{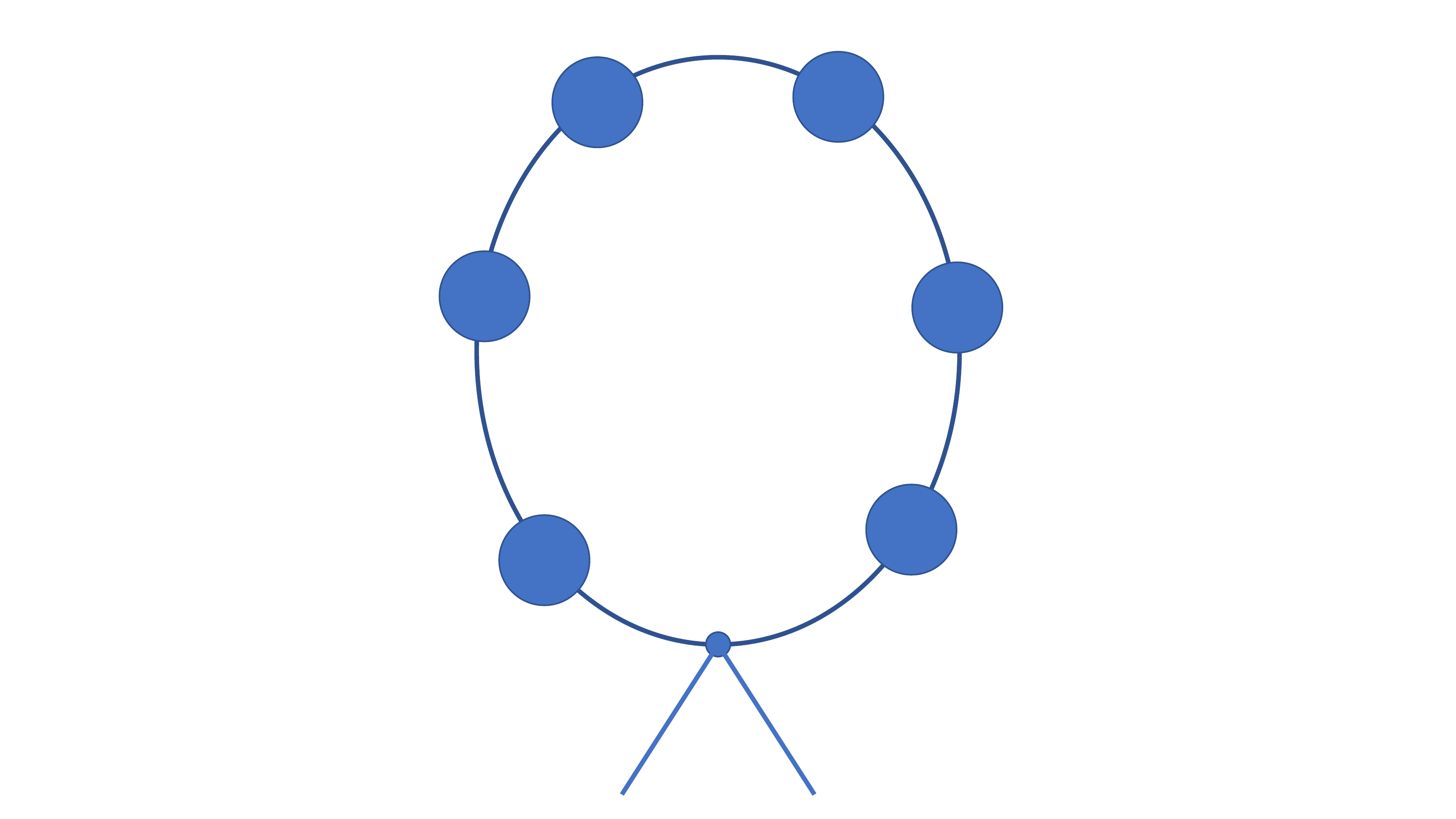}
\caption{\label{gtad1}
A generalized tadpole. Each blue dot corresponds to a 2PI biped.}
\end{figure}

In this section we study the upper bound for the counter-term $\delta\mu(\lambda)$. 
Three different terms may  contribute to $\delta\mu(\lambda)$: the tadpole term, the integration of the self-energy function:
\be
\Sigma^{\le r_{max}}(x,\l)=\sum_{r=0}^{r_{max}}\int d^3y\ \Sigma^r(x, y,\l),
\ee
and the {\it the generalized tadpole term}, which is a tadpole term whose internal lines are decorated by 1PI bipeds. See Figure \ref{gtad1} for an illustration.
First of all, by Lemma \ref{tadmain1}, the amplitude of a tadpole is bounded by:
\be
\Vert T\Vert_{L^\infty}\le C|\lambda|,\nn
\ee
satisfying the bound in Theorem \ref{flowmu}. Secondly, we have:
\be\label{localmu1}
\sup_x|\Sigma^{\le r_{max}}(x,\l)|\le \sum_{r=0}^{r_{max}}\int\ d^3y\ \sup_x|\Sigma^r(x, y,\l)|,
\ee
in which the integrand is bounded by (cf. Theorem \ref{maina})
\be
 C_1|\lambda|^2 r^2 \sup_x\sup_{\sigma\in\sigma_{\cG^r}}\g^{-3r(\sigma)}e^{-cd^\alpha_\sigma(x,y)},\nn
\ee
for some positive constant $C_1$ independent of $T$ and $\l$. Perform the integration in \eqref{localmu1} along the spanning tree in the 1PI graph, the spatial integration is bounded by 
\be|\int d^3y e^{-cd^\alpha_\sigma(x,y)}|\le C_2.\gamma^{2r_\cT},
\ee
where $r_\cT$ is the maximal scale index among the tree propagators. Combine the above two terms, we find that, these exists another positive constant $C_3$ such that
\be\label{local3}
\sum_r\int\ dy_0 dy^{+}dy^{-}\ |\Sigma^r(x, y,\l)|\le C_3 |\lambda|^2,
\ee
which is bounded by $C_3|\lambda|$ for $\lambda\in\RR_T$. Now we consider the amplitude of a generalized tadpole, which is formed by contracting a chain of bipeds with a bare vertex. Let $T^g_n$ be a generalized tadpole which contain $n$ irreducible and renormalized bipeds and $n+1$ propagators connecting these bipeds (See Figure \ref{gtad1} for an illustration). Let the scale index of an external propagators be $r^e$ and the lowest scale index of the propagators in the biped be $r^i$, the amplitude of a generalized tadpole is
\be T^\Sigma(r^e,r^i,\lambda,x)= \int d{x_1}\int dy_1 C^{r^e}(x,x_1)(1-\tau)\Sigma^{r^i}(x_1,y_1,\l),
\ee
and we have
\bea
&&\Vert (1-\tau)\Sigma^{r^i}(x_1,y_1,\l)\Vert_{L^\infty}\nn\\
&&\le\sup_{x}\int dx_1 dy_1|x_{1,0}-y_{1,0}|\cdot\vert\Sigma^{r^i}(x_1,y_1,\l)\vert\cdot|
\frac{\partial}{\partial{x_{1,0}}} C^{r^e}(x,x_1)|\nn\\
&&+\sup_{x}
\int dx_1 dy_1|x_1^{+}-y_1^{+}|\cdot|x_1^{-}-y_1^{-}|\cdot\vert\Sigma^{r^i}(x_1,y_1,\l)\vert\cdot|
\frac{\partial}{\partial{x_1^{+}}}\frac{\partial}{\partial{x_1^{-}}} C^{r^e}(x,x_1)|.\nn
\eea
Using the fact that $j^e+s_{+}^e+s_{-}^e=2r^e$ and $s_{+}^e+s_{-}^e\ge 6r^e/5-j_0/5\ge r^e$, for $r^e\ge j_0$, we find that there exist positive constants $C_4$ and $C_5$ such that
\be\sup_{x}|\frac{\partial}{\partial{x_{1,0}}} C^{r^e}(x,x_1)|\le C_4\g^{-2r^e}e^{-d(x,x_1)},\ 
\sup_{x}|\frac{\partial}{\partial{x_1^{+}}}\frac{\partial}{\partial{x_1^{-}}}C^{r^e}(x,x_1)|\le C_5\g^{-2r^e}e^{-d(x,x_1)}.\nn\ee
By Theorem \ref{maina}, Formula \eqref{x2pt} and \eqref{x2ptc}, 
we find that, after performing the integration, each of the above term is bounded by 
$$\max\{C_4,C_5\}\g^{2r^e}\g^{2r^i}\g^{-2r^e}\g^{-2r^i}\le \max\{C_4,C_5\}.$$

Notice that there is a propagator in $T^g_n$ whose position coordinates are not integrated and using Lemma \ref{secbi}, we have
\be
\Vert T^g_n\Vert_{L^\infty}\le\prod_{i=1}^n\Vert\sum_{r^i}  T^\Sigma_i(r^e,r^i,\lambda,x)\Vert_{L^\infty}\cdot\Vert C^{r^e}(x_n,x)\Vert_{L^\infty}\le K^n |\lambda|^{2n+1} |\log \mu_0T|^n\g^{-r^e}.
\ee
Summing over the indices $r^e$ and using the fact that $|\lambda\log^2 \mu_0T|<c$, we can easily see that there exists a positive constant $C_6$, independent of $T$ and $\l$, such that
\be
\sum_{n=0}^\infty |T^g_n|\le 2 C_6 |\lambda|.
\ee

Summing up all the local terms and let $C'=C_1+C_3+2C_6$, then the amplitudes of all the local terms are bounded by $C'|\lambda|$. Hence we proved Theorem \ref{flowmu}. 


\section{Conclusions and Perspectives}
In this paper we constructed the renormalized $2$-point Schwinger functions and the self-energy functions in the $2d$ square-Hubbard model with renormalized chemical potential $\mu$ slightly smaller than $2$, in which the Fermi surface is close to a square and von Hove singularities are not present. We have established the optimal upper bounds for the self-energy function as well as its second derivative. Our result suggest that the ground state is not a Fermi liquid in the mathematical precise sense of Salmhofer. This work is an important step towards the study of the cross-over phenomenon of the Fermi liquid and non-Fermi liquid behaviors in the Hubbard model on the square lattice. In a future paper we will study the lower bounds for the self-energy as well as its second derivatives in this model. The similarities between the square-Hubbard model and the honeycomb-Hubbard model suggest universal properties in the 2-d Hubbard model, which also deserve further investigations. 


\section{Appendix}
In this part we consider the geometry of the Fermi surface. Consider a shell $$\cD_j=\{(k_+,k_-)\in\cB:(2\cos \frac{k_+}{2}\cos \frac{k_-}{2}-\mu_0)^2\le\g^{-2j},\ \g^{-j}<\mu_0\}$$ with fixed $\mu_0$. The boundaries of $\cD_j$ are
\bea\label{curvj}
\cF_0^{(j),+}(\mu_0)&=&\{(k_+,k_-)\in\cD_j:2\cos \frac{k_+}{2}\cos \frac{k_-}{2}=\mu_0-\g^{-j}, |k_\pm|\ge |P_{\cF_0}(k_\pm)|\ \},\nn\\
\cF_0^{(j),-}(\mu_0)&=&\{(k_+,k_-)\in\cD_j:2\cos \frac{k_+}{2}\cos \frac{k_-}{2}=\mu_0+\g^{-j}, |k_\pm|\le |P_{\cF_0}(k_\pm)|\ \}.\nn
\eea 
\begin{proposition}\label{geom}
The curvature radius of the curves $\cF_0^{(j),\pm}$ are given by
\be
R^{(j),\pm}(k_+,k_-,\mu_0)=\frac{4\big[\cos^2\frac{k_+}{2}\sin^2 \frac{k_-}{2}+\sin^2\frac{k_+}{2}\cos^2\frac{k_-}{2}\big]^{3/2}}{(\mu_0\mp\g^{-j})(\sin^2\frac{k_+}{2}+\sin^2\frac{k_-}{2})},
\ee
respectively, which have several monotonic intervals on $\cF_0^{(j),\pm}$. Eg. for any fixed
$k_-\in\big[\ 2\arccos[(\mu_0\pm\g^{-j})/2]^{1/2}, 2\arccos(\mu_0\pm\g^{-j})/2\ \big]$, $R^{(j),\pm}(k_+,k_-)$ are decreasing functions in $k_+$, and reach their maximal values $R_{max}=\frac{4[1-(\mu_0\pm\g^{-j})^2/4]^{1/2}}{(\mu_0\pm\g^{-j})}$ at the point $(0,2\arccos(\mu_0\pm\g^{-j})/2)$, and reach the minimum values $R_{min}=2[\mu_0\pm\g^{-j}]^{1/2}[1-(\mu_0\pm\g^{-j})^2/4]^{1/2}$ at $(2\arccos(\mu_0\pm\g^{-j})/2,2\arccos[(\mu_0\pm\g^{-j})/2]^{1/2})$. 
\end{proposition}
This proposition can be proved by elementary geometry. Other monotonic intervals of $R^{(j),\pm}$ can be obtained using symmetry of $\cF_0^{(j),\pm}$.   When $j\rightarrow\infty$, we obtain the curvature radius of $\cF_0$.
\begin{remark}
Setting $\mu_0=\g^{-j}$ for fixed $j\ge1$, we have $R_{max}^{(j),+}(k_+,k_-)=\infty$ and $R_{min}^{(j),+}(k_+,k_-)=0$, so that $\cF_0^{(j),+}$ becomes a square. Taking the limit $j\rightarrow\infty$ while keeping $\mu_0=\g^{-j}$, $\{\cF_0^{(j),+}(\mu_0)\}$ becomes the Fermi surface at half-filling \cite{Riv04}.
\end{remark}
\begin{definition}\label{fl}
Define a set of curves $\cF_0^{(j),+,flat}=\cF_0^{(j),+}(\g^{-j}), j=0,1,2,\cdots,j_{max}$, and\\
$\cF_0^{flat}=\lim_{j\rightarrow\infty,\m_0=\g^{-j}}\cF_0^{(j),+}(\mu_0).$
$\cF_0^{flat}$ is called the half-filing limit of the curves $\{\cF_0^{(j),+}(\mu_0)\}$.
\end{definition}
Let $\delta^j$ be the distance of the curve $\cF_0^j$ from $\cF_0$, we have:
\be
\delta^j=\frac{\g^{-j}}{2(\cos^2 \frac{k_+}{2}-\frac{\mu_0^2}{4})^{\frac12}+2[\cos^2 \frac{k_+}{2}-\frac{\mu_0^2}{4}+\mu_0\gamma^{-j}]^{\frac12}}.\nn
\ee
Now consider a chord whose distance to the boundary of the shell is $\delta$ and whose length is $w$, we have $\g^{-j}\le \delta\le \frac{\g^{-j}}{\sqrt\mu_0}$. Using the relation $w=\sqrt{R\delta}$, we have $\g^{-\frac{j}{2}}\le w\le\frac{\g^{-\frac{j}{2}}}{\sqrt\mu}$. So that the sectors at each scale $j$ can be defined as the rectangles parallel to the axis $k_+=0$ and $k_-=0$, whose sides along the $k_+$ direction (called the length) are not longer than $w$ and whose sides along the $k_-$ direction (called the height) are not longer than $d$, which lead to Definition \ref{defsec}.
\medskip
\noindent{\bf Acknowledgments}
The author is grateful to V. Rivasseau for useful comments on the manuscript. This work is supported by NSFC No.12071099.

\end{document}